%% file: main.tex
\documentclass[a4paper, USenglish, autoref, thm-restate]{lipics-v2021}

\pdfoutput=1 %

\input{preamble}

\input{notation-macros}

\title{\framework{}: Lagom Multi-Query Sketch for High-Rate Online Analytics}

\author{Martin Hilgendorf}{Chalmers University of Technology and University of Gothenburg, Sweden}{martin.hilgendorf@chalmers.se}{https://orcid.org/0009-0008-6333-3503}{}

\author{Marina Papatriantafilou}{Chalmers University of Technology and University of Gothenburg, Sweden}{ptrianta@chalmers.se}{https://orcid.org/0000-0001-9094-8871}{}

\authorrunning{M. Hilgendorf, and M. Papatriantafilou} %

\Copyright{Martin Hilgendorf and Marina Papatriantafilou} %



\supplement{The source code for the implementation of \framework{} can be found here:}
\supplementdetails{Software}{https://gitlab.com/disc2025-380/lmq-sketch}

\funding{Work supported by the \href{https://www.vr.se/}{Swedish Research Council} project EPITOME 2021-05424, by the \href{https://marie-sklodowska-curie-actions.ec.europa.eu/}{Marie Skłodowska-Curie Doctoral Network} \href{www.relax-dn.eu}{RELAX-DN}, funded by the European Union under Horizon Europe 2021-2027 Framework Programme (grant nr. 101072456) and by Chalmers University AoA frameworks Energy and Production, WP. INDEED, and `Scalability, Big Data and AI'.}  %

\EventEditors{Dariusz Kowalski}
\EventNoEds{1}
\EventLongTitle{39th International Symposium on Distributed Computing (DISC 2025)}
\EventShortTitle{DISC 2025}
\EventAcronym{DISC}
\EventYear{2025}
\EventDate{October 27 -- October 31, 2025.}
\EventLocation{Berlin, Germany}
\EventLogo{}
\SeriesVolume{XX}
\ArticleNo{YY}

\begin{document}

    \maketitle

    \begin{abstract}
        \input{include/abstract.tex}

    \end{abstract}

    \newpage

    \input{include/01_introduction}

    \input{include/02_prel}

    \input{include/03_problem}

    \input{include/04_overview}

    \input{include/05_analysis}

    \input{include/06_evaluation}

    \input{include/07_relatedwork}

    \input{include/08_conclusions}

    \bibliography{refs-concurrency,refs-datasets,refs-sketches,refs-software}

    \appendix
    \input{include/a_proofs}

    \input{include/b_evaluation}
\end{document}

%% file: preamble.tex
\usepackage{siunitx}
\usepackage[linesnumbered,vlined,ruled,algonl]{algorithm2e}
\input{algos/common.tex}

\usepackage{booktabs}

\usepackage{makecell}

\renewcommand\theadfont{\bfseries}

\usepackage{hyperref}

\usepackage[normalem]{ulem}

\usepackage{subcaption}

\usepackage{svg}

\svgpath{ {figures/} }
\graphicspath{ {figures/} }

\DeclareUnicodeCharacter{2212}{\textendash}

\usepackage{tikz}
\usetikzlibrary{patterns,patterns.meta,tikzmark}

\tikzset{every picture/.style={/utils/exec={\small}}}

\usepackage[at]{easylist}
\NewList(%
Hang=true,Space=0pt,Space*=0pt,Hide=1000,%
Margin1=1.5em,Style1*=\textbullet\hskip .5em,%
Margin2=3.7em,Style2*=--\hskip .5em,%
Margin3=5.9em,Style3*=$\ast$\hskip .5em,%
Margin4=7.8em,Style4*=$\cdot$\hskip .5em)

\newcommand{\algoHighlight}[4]{%
    \tikz[remember picture, overlay]{%
        \fill[#2!50, opacity=0.5, blend mode=darken]%
        ([xshift={#3[0]}, yshift={#3[1]}]pic cs:#1:start) rectangle ([xshift={#4[0]}, yshift={#4[1]}]pic cs:#1:end);%
    }%
}

\newcommand{\algoHighlightSingleLine}[4]{%
    \protect\algoHighlight{#1}{#2}{{#3,7pt}}{{#4,-2pt}}%
}

\newcommand{\algobox}[1]{%
\noindent\fcolorbox{gray}{white}{%
    \parbox{\dimexpr\linewidth - 2\fboxsep}{%
        \phantomsection\small#1%
    }%
}%
}

\expandafter\def\expandafter\normalsize\expandafter{%
    \normalsize%
    \setlength\abovedisplayskip{3pt}%
    \setlength\belowdisplayskip{1pt}%
    \setlength\abovedisplayshortskip{-4pt}%
    \setlength\belowdisplayshortskip{1pt}%
}

\nolinenumbers
\hideLIPIcs

%% file: algos/common.tex
\SetKwProg{Fn}{Function}{}{end}

\SetKwFunction{findOwnerFn}{findOwner}
\SetKwFunction{cmsInsertFn}{update}
\SetKwFunction{cmsInsertEnhFn}{update\tikzmark{tm:asketch:insert-enhanced:update-enhanced:start}Enhanced\tikzmark{tm:asketch:insert-enhanced:update-enhanced:end}}
\SetKwFunction{cmsInsertAndPQFn}{updateAndPQ}
\SetKwFunction{cmsInsertAndPQEnhFn}{updateAndPQ\tikzmark{tm:asketch:insert-enhanced:update-and-pq-enhanced:start}Enhanced\tikzmark{tm:asketch:insert-enhanced:update-and-pq-enhanced:end}}
\SetKwFunction{asketchInsertFn}{\asketchInsert}
\SetKwFunction{asketchQueryFn}{\asketchQuery}
\SetKwFunction{asketchInsertEnhancedFn}{ASketchUpdate\tikzmark{tm:lagom:ppi:asketch-update-enhanced:start}Enhanced\tikzmark{tm:lagom:ppi:asketch-update-enhanced:end}}
\SetKwFunction{insertAlgoFn}{\insertAlgo}
\SetKwFunction{processPendingInsertsFn}{\ppi}
\SetKwFunction{pointQueryFn}{pointQuery}
\SetKwFunction{queryFOneFn}{queryF1}
\SetKwFunction{queryFTwoNosyncFn}{queryF2Nosync}
\SetKwFunction{queryFTwoLagomFn}{queryF2Lagom}

\SetKwComment{Comment}{$\triangleright$\,}{}

\SetNlSty{bfseries}{\color{black}}{}

\SetAlFnt{\scriptsize}

\SetFuncArgSty{textnormal}  % In function arguments
\SetArgSty{textnormal}      % In if/loop conditions

\SetInd{0.2em}{0.8em}

\SetVlineSkip{1pt}

\SetAlgoSkip{}

%% file: notation-macros.tex
\DeclareMathOperator*{\argmin}{argmin}
\newcommand{\pluseq}{\ensuremath{\mathbin{{+}{=}}}}
\newcommand{\plusplus}{\ensuremath{\mathord{{++}}}}

\newcommand{\Cpp}{C\nolinebreak[4]\hspace{-.05em}\raisebox{.2ex}{\small\bf ++}}

\newcommand{\epsdelta}{\ensuremath{(\epsilon,\delta)}}

\newcommand{\pq}{\ensuremath{PQ}}
\newcommand{\fone}{\ensuremath{F_1}}
\newcommand{\ftwo}{\ensuremath{F_2}}
\newcommand{\textpq}{\textit{PQ}}
\newcommand{\textfone}{\texorpdfstring{\textit{F\textsubscript{1}}}{F1}}
\newcommand{\textftwo}{\texorpdfstring{\textit{F\textsubscript{2}}}{F2}}
\newcommand{\foneest}{\ensuremath{\hat{\fone}}}
\newcommand{\fonenosync}{\ensuremath{\smash{\foneest{}^{\nosync}}}}
\newcommand{\ftwoest}{\ensuremath{\smash{\hat{\ftwo}}}}
\newcommand{\ftwoestm}[1]{\ensuremath{\smash{\hat{\ftwo}{}^{#1}}}}

\newcommand{\heavyKeyFTwoNosync}{\ensuremath{\mathcal{H}^{\nosync}}}
\newcommand{\heavyKeyFTwoLagom}{\ensuremath{\mathcal{H}^{\lagom}}}

\newcommand{\cmsName}{\ensuremath{\text{\textsf{CMS}}}}
\newcommand{\cmsIndexed}[2]{\ensuremath{\cmsName[#1, #2]}}
\newcommand{\sketchWidth}{\ensuremath{K}}
\newcommand{\sketchHeight}{\ensuremath{H}}
\newcommand{\sketchRowIndex}{\ensuremath{j}}
\newcommand{\sketchColIndex}{\ensuremath{k}}

\newcommand{\hashRow}[1]{\ensuremath{h_#1}}

\newcommand{\dataKey}{\ensuremath{a}}
\newcommand{\dataValue}{\ensuremath{v}}
\newcommand{\dataTuple}{\ensuremath{(\dataKey{}, \dataValue{})}}

\newcommand{\frequency}[1]{\ensuremath{f(#1)}}
\newcommand{\frequencyEst}[1]{\ensuremath{\hat{f}(#1)}}

\newcommand{\delThreads}{\ensuremath{P}}  %
\newcommand{\delFilterSlots}{\ensuremath{C}}
\newcommand{\delFilterMaxCount}{\ensuremath{B}}

\newcommand{\delThreadIIndex}{\ensuremath{i}}
\newcommand{\delThreadIIIndex}{\ensuremath{j}}

\newcommand{\delThreadName}{T}
\newcommand{\delThreadI}[1]{\ensuremath{\smash{\delThreadName_{#1}}}}

\newcommand{\sketchName}{\ensuremath{\mathsf{Sk}}}
\newcommand{\cmsPartialFtwo}{F2partials}

\newcommand{\delOwner}[1]{\ensuremath{\mathsf{Owner}(#1)}}

\newcommand{\augFilterName}{\ensuremath{\mathsf{AF}}}
\newcommand{\augFilterCount}[1]{\ensuremath{\augFilterName[#1]}}
\newcommand{\augFilterOldCount}[1]{\ensuremath{\smash{\augFilterName^{\text{old}}[#1]}}}
\newcommand{\augFilterProjection}[1]{\ensuremath{\smash{\augFilterName^{\text{mavg}}[#1]}}}
\newcommand{\augFilterDrift}{\ensuremath{\Delta}}

\newcommand{\delFilterName}{\ensuremath{\mathsf{DF}}}
\newcommand{\delFilterTarget}[1]{\ensuremath{\smash{\delFilterName_{#1}}}}
\newcommand{\delFilterCount}[2]{\ensuremath{\smash{\delFilterName_{#1}[#2]}}}
\newcommand{\delFilterFilledSlots}{size}
\newcommand{\delFilterBufferedCount}{sum}

\newcommand{\delThreadSketch}[1]{\ensuremath{\delThreadI{#1}.\sketchName{}}}
\newcommand{\delThreadAugFilter}[1]{\ensuremath{\delThreadI{#1}.\augFilterName{}}}
\newcommand{\delThreadAugFilterCount}[2]{\ensuremath{\delThreadI{#1}.\augFilterCount{#2}}}
\newcommand{\delThreadAugFilterOldCount}[2]{\ensuremath{\delThreadI{#1}.\augFilterOldCount{#2}}}
\newcommand{\delThreadAugFilterProjection}[2]{\ensuremath{\delThreadI{#1}.\augFilterProjection{#2}}}

\newcommand{\delThreadDelFilter}[2]{\ensuremath{\delThreadI{#1}.\delFilterName_{#2}}}
\newcommand{\delThreadDelFilterCount}[3]{\ensuremath{\delThreadI{#1}.\delFilterCount{#2}{#3}}}

\newcommand{\delThreadFullFilters}{pendingFilters}
\newcommand{\delThreadInsertedElements}{F1partial}
\newcommand{\delThreadFlag}{beingScanned}
\newcommand{\versionOne}{\ensuremath{\mathsf{V1}}}
\newcommand{\versionTwo}{\ensuremath{\mathsf{V2}}}
\newcommand{\delThreadTSone}[1]{\ensuremath{\delThreadI{#1}.\versionOne{}}}
\newcommand{\delThreadTStwo}[1]{\ensuremath{\delThreadI{#1}.\versionTwo{}}}

\newcommand{\asketchInsert}{ASketchUpdate}
\newcommand{\asketchQuery}{ASketchQuery}
\newcommand{\insertAlgo}{update}
\newcommand{\ppi}{processDelegatedUpdates}

\newcommand{\cms}{\texorpdfstring{Count\nobreakdashes-Min Sketch}{Count-Min Sketch}}
\newcommand{\asketch}{Augmented Sketch}
\newcommand{\delsketch}{Delegation Sketch}
\newcommand{\swskt}{SW\nobreakdash-SKT}
\newcommand{\fastagms}{Fast\nobreakdashes-AGMS}
\newcommand{\cmm}{\emph{CM\textsuperscript{--}}}
\newcommand{\cmp}{\emph{\texorpdfstring{CM\textsuperscript{+}}{CM+}}}

\newcommand{\fwshort}{LMQ}
\newcommand{\framework}{\texorpdfstring{LMQ\nobreakdashes-Sketch}{LMQ-Sketch}}
\newcommand{\fullsync}{\textsc{Strict}}  %
\newcommand{\nosync}{\textsc{Nosync}}  %
\newcommand{\lagom}{\textsc{Lagom}}  %

\newcommand{\opstart}[1]{\ensuremath{#1^{\mathit{start}}}}
\newcommand{\opend}[1]{\ensuremath{#1^{\mathit{end}}}}

\newcommand{\qop}{\ensuremath{Q}}
\newcommand{\qstart}{\smash{\opstart{\qop}}}
\newcommand{\qend}{\smash{\opend{\qop}}}
\newcommand{\precedes}{\ensuremath{\longrightarrow}}

\newcommand{\subop}[2]{\ensuremath{\smash{#1\big\lvert_{#2}}}}

\newcommand{\analysisWide}{\textsc{Wide}}
\newcommand{\analysisWideConc}{\emph{\ensuremath{\text{CM}^{\text{+}}_{\text{conc}}}}}

\newcommand{\analysisPart}{\textsc{PartCMS}}
\newcommand{\analysisPartSeq}{\emph{\analysisPart{}\textsuperscript{+}}}
\newcommand{\analysisPartConc}{\emph{\ensuremath{\text{\analysisPart{}}^{\text{+}}_{\text{conc}}}}}

\newcommand{\analysisPartASketch}{\textsc{PartAS}}
\newcommand{\analysisPartASketchSeq}{\emph{\analysisPartASketch{}\textsuperscript{+}}}
\newcommand{\analysisPartASketchConc}{\emph{\ensuremath{\text{\analysisPartASketch{}}^{\text{+}}_{\text{conc}}}}}

\newcommand{\analysisOurs}{\framework{}}
\newcommand{\analysisOursFullSeq}{\emph{\fwshort{}\nobreakdashes-all\textsuperscript{+}}}
\newcommand{\analysisOursFullConc}{\ftwoestm{\nosync}}
\newcommand{\analysisOursRelaxedSeq}{\analysisPartASketchSeq{}}
\newcommand{\analysisOursRelaxedConc}{\analysisPartASketchConc{}}
\newcommand{\analysisOursProjSeq}{\emph{\fwshort{}\nobreakdashes-proj\textsuperscript{+}}}
\newcommand{\analysisOursProjConc}{\ftwoestm{\lagom}}

%% file: include/abstract.tex
Data sketches balance resource efficiency with controllable approximations for extracting features in high-volume, high-rate data.
Two important points of interest are highlighted separately in recent works; namely, to
(1)~answer multiple types of queries from one pass, and
(2)~query concurrently with updates.
Several fundamental challenges arise when integrating these directions, which we tackle in this work.

We investigate the trade-offs to be balanced and synthesize key ideas into
\framework{}, a single, composite data sketch supporting multiple queries (frequency point queries, frequency moments \textfone{}, and \textftwo{}) concurrently with updates.
Our method~`\lagom{}' is a cornerstone of~\framework{} for low-latency global querying (<\qty{100}{\micro\second}), combining freshness, timeliness, and accuracy with a low memory footprint and high throughput (>2B updates/s).
We analyze and evaluate the accuracy of \lagom{}, which builds on a simple geometric argument and efficiently combines work distribution with synchronization for proper concurrency semantics -- \emph{monotonicity of operations} and \emph{intermediate value linearizability}.
Comparing with state-of-the-art methods (which, as mentioned, only cover either mixed queries or concurrency),
\framework{} shows highly competitive throughput, with additional accuracy guarantees and concurrency semantics, while also reducing the required memory budget by an order of magnitude.
We expect the methodology to have broader impact on concurrent multi-query sketches.

%% file: include/01_introduction.tex
\section{Introduction}
\label{sec:introduction}

Data sketches provide approximate summaries of data streams
and can answer questions of interest efficiently, with bounded memory requirements.
Examples include estimation of element frequencies, set/multi-set size, frequency moments (norms), frequent elements, distributions, quantiles, and more~\cite{alonTrackingJoinSelfjoin1999, buragohainQuantilesStreams2009, charikarFindingFrequentItems2002, cormodeSketchingStreamsNet2005, cormodeImprovedDataStream2005, cormodeSmallSummariesBig2020}.
Summarizations involve suitable hash functions, and the results are controllable approximations of the targeted aggregate -- commonly for skewed data -- in form of \epsdelta{} guarantees (the estimate deviates at most $\epsilon$, with probability at least $1-\delta$).
The approximation reflects a \emph{trade-off between accuracy and sketch size}, and, in consequence, the operations' time cost.
Due to their usefulness in data analytics and feature extraction, sketches get a major role in  data processing platforms~\cite{ApacheDruidApache, DataSketchesApacheDataSketches, haberCountMinSketchArt2022, koganTdigestNewProbabilistic2023, noremProbabilisticDataStructures2022}.

\noindent\textbf{\textsf{Mixed query sketches~}}
Data analysis requires more information than a single metric.
Typically, this would require one sketch for each metric to be tracked~\cite{chiosaSKTOnepassMultisketch2021},
requiring additional memory and processing, a prohibitive overhead, stressed as early as in~\cite{cormodeImprovedDataStream2005}.
Can it be done differently? Important questions regarding having a \emph{single sketch} for answering \emph{mixed queries} have also been asked in works such as~\cite{dobraSketchBasedMultiQueryProcessing2016, manousisEnablingEfficientGeneral2022, punterOmniSketchEfficientMultiDimensional2023}, showing potential to improve approximation guarantees and quality of results compared to using multiple separate sketches.
A notable sketch supporting, with auxiliary data structures, multiple statistics (element frequency, frequency moments, frequent elements, quantiles and more) is the \cms{} (CMS)~\cite{cormodeImprovedDataStream2005, cormodeSmallSummariesBig2020}.
Besides, at the core of many statistics are \emph{frequency moments}
(the $i$-th such defined as the sum of the $i$-th power of the frequency of each element);
e.g., \ftwo{}, also called `surprise number', has uses in linear algebra~\cite{woodruffSketchingToolNumerical2014}, calculating the Gini index~\cite{cerianiOriginsGiniIndex2012, giniVariabilitaMutabilita1912}, evaluating skewness of distributions, quantiles~\cite{gilbertHowSummarizeUniverse2002}, or wavelet synopses helping for compression~\cite{cormodeFastApproximateWavelet2006, garofalakisDiscreteWaveletTransform2009,  gilbertOnepassWaveletDecompositions2003}.
In conjunction with other metrics, \ftwo{} enables to detect anomalies~\cite{cormodeSketchingStreamsNet2005, krishnamurthySketchbasedChangeDetection2003, matosGenericChangeDetection2022, tongSketchAccelerationFPGA2018}
and is fundamental for universal sketching~\cite{bravermanZerooneFrequencyLaws2010, liuOneSketchRule2016}.
\ftwo{} can be estimated via, e.g., CMS~\cite{cormodeSummarizingMiningSkewed2005} or \fastagms{}~\cite{cormodeSketchingStreamsNet2005}, an efficient successor of the pioneer AMS sketch~\cite{alonSpaceComplexityApproximating1999}.\looseness=-1

\noindent\textbf{\textsf{Concurrency~}}
is both a \emph{necessity} given the rates of real-world streams, and a \emph{challenge}, since, on one hand, the perceived order of updates by queries influences accuracy, and, on the other hand, synchronization overhead can dramatically influence operation timeliness and result freshness.
Work to address concurrency among queries and updates on sketching was initiated in the recent years, for single-query sketches~\cite{eliaszadaQuancurrentConcurrentQuantiles2023, rinbergIntermediateValueLinearizability2023, rinbergFastConcurrentData2022, stylianopoulosDelegationSketchParallel2020}.

\noindent\textbf{\textsf{Our Targets~}}
We integrate these two directions, %
opening up the challenge to balance accuracy with resource footprint and timeliness, through workload distribution, data structure design, and synchronization for proper consistency guarantees.
We target high-rate sketching for \fone{}, \ftwo{}, and individual element frequencies (point queries) on a single, low memory-footprint composite data structure with concurrent updates and queries.

\noindent\textbf{\textsf{Challenges~}}
Concurrency adds new factors and trade-offs to consider.
The influence is complex: stronger consistency can imply better compliance with sequential sketch accuracy bounds but higher synchronization overhead, thus increasing query latency, which can aggravate accuracy due to staleness.
\emph{Intermediate Value Linearizability} (IVL)~\cite{rinbergIntermediateValueLinearizability2023} is a modular consistency property
suitable for sketches, admitting more efficient implementations compared to requiring linearizability (which is unnecessarily strong as sketches approximate by definition) and
can preserve \epsdelta{} bounds of the sequential counterparts.
An additional challenge regarding concurrent use of separate sketches for mixed queries is that they can be inconsistent.
In a nutshell, various
trade-offs surround the metrics of interest: set of queries, result quality (accuracy, consistency), resource footprint, operation timeliness, and scalability.\looseness=-1

\noindent\textbf{\textsf{Idea and Contributions~}}
We study these trade-offs and construct \emph{\framework{}} (\emph{Lagom\footnotemark{} Multi-Query Sketch}), a low memory-footprint sketch supporting concurrent and mixed frequency queries.
We partition the input domain and data structure, and delegate work among threads, a design also present in~\cite{stylianopoulosDelegationSketchParallel2020} albeit for point queries only, with benefits in accuracy and memory-efficiency.
When answering \emph{global queries} that span multiple partitions, e.g. \fone{} and \ftwo{},
increased scan latency and synchronization overhead can be detrimental to both timeliness and the quality of query results.
Our IVL-aligned algorithmic design \lagom{} demonstrates significant gains in efficiency by combining partitioning with lightweight synchronization and concurrency-aware helping;
it gathers `just enough' (shown using a geometric argument) information from threads' states using partial results, lowering query complexity yet maintaining accuracy in line with sequential methods.
We also show cross-query consistency properties of \framework{}, in particular about \emph{monotonicity of scans}~\cite{dworkTimeLapseSnapshots1999}.\looseness=-1

\footnotetext{Lagom (\href{https://en.wikipedia.org/wiki/Lagom}{\textsf{link}}) describes `just the right amount', indicating appropriateness rather than suggesting lack.\looseness=-1}

The analysis is complemented by a detailed empirical study relative to state-of-the-art methods: \swskt{}~\cite{chiosaSKTOnepassMultisketch2021}, employing separate sketches for
multiple queries albeit non-concurrently, and \delsketch{}~\cite{stylianopoulosDelegationSketchParallel2020}, supporting concurrency but for point queries only.
We define elementary baselines based on the competing factors in the trade-offs.
Our evaluation shows how \framework{} efficiently balances memory, accuracy, and concurrency,
scaling beyond 2B~updates/sec concurrently with high-rate queries.
\ftwo{} query latency between \qtyrange[range-phrase=--, range-units=single]{1}{100}{\micro\second} implies freshness with error below \qty{0.01}{\percent}, with a sketch just \qty{4}{\mebi\byte} in size.

\noindent\textbf{\textsf{Roadmap~}}
After preliminaries (§~\ref{sec:background}) and problem analysis (§~\ref{sec:problem}), we present and analyze the methods constituting \framework{} (§~\ref{sec:overview}, \ref{sec:analysis}), detail our empirical evaluation (§~\ref{sec:eval}, with open-source implementation~\cite{anonymousauthorsLMQSketch}) and other related work (§~\ref{sec:relatedwork}), concluding in §~\ref{sec:conclusion}.
Throughout, claims are backed by the main ideas; proofs or proof sketches are provided in \autoref{appendix:proofs}.

%% file: include/02_prel.tex
\section{Background}  \label{sec:background}

\noindent\textbf{\textsf{\cms{} (CMS) and Frequency Moments~}}
CMS~\cite{cormodeImprovedDataStream2005} estimates the frequency of keys in the input stream, via \emph{point queries}, with potential overestimation by a factor $\epsilon$ with probability $1 - \delta$.
It uses a matrix of counters, \cmsName{}[$\sketchHeight{}\!\times\!\sketchWidth{}$], alongside \sketchHeight{} hash functions, each mapped to a row.
\sketchHeight{} and \sketchWidth{} balance memory against accuracy,
by $\sketchHeight = \left\lceil e / \epsilon \right\rceil$ and $\sketchWidth = \left\lceil \ln(1/\delta) \right\rceil$. %
A CMS update for a key \dataKey{} increments $\cmsIndexed{\sketchRowIndex}{\hashRow{\sketchRowIndex{}}(\dataKey{})}$ for each row \sketchRowIndex{}.
To answer a point query for \dataKey{}, its frequency \frequency{\dataKey} is estimated as
    $\frequencyEst{\dataKey} = \min_{\sketchRowIndex{}} \cmsIndexed{\sketchRowIndex}{\hashRow{\sketchRowIndex{}}(\dataKey{})}$.

For a stream $S$ of keys from a domain $U$,
the $n$-th frequency moment is~\cite{woodruffFrequencyMoments2018}:
   $F_n = \sum_{\dataKey{} \in U} \frequency{\dataKey{}}^n$.
CMS provides an exact \fone{} as $\sum_{\sketchColIndex}^{\sketchWidth}\!\cmsIndexed{\sketchRowIndex}{\sketchColIndex}$ for any row \sketchRowIndex{}.
To estimate \ftwo{}, which gives an indication of the skewness of the frequency distribution, \cite{cormodeSummarizingMiningSkewed2005} introduces \cmm{} and \cmp{} for different levels of skewness, encoded by the $z$ parameter of a Zipfian, as many real-world phenomena follow Zipf's Law;
\cmm{} for $z \leq 1$, while \cmp{} yields better accuracy for skewed data ($z > 1$).
Similar to CMS point queries, \cmp{} can only overestimate.
\begin{definition}[\cmp{} \cite{cormodeSummarizingMiningSkewed2005}]  \label{defn:cm+}
    \cmp{} estimates \ftwo{} from a given \textup{\cmsName{}} by $\min_{\sketchRowIndex{}} \sum_{\sketchColIndex}^{\sketchWidth} \textup{\cmsIndexed{\sketchRowIndex}{\sketchColIndex}}^2$ with relative error $1 + \epsilon$ with probability $1 - \delta = 1 - \smash{\frac{3}{4}^{-H}}$, where $\epsilon = O\!\left(K^{\frac{-(1+z)}{2}}\right)$.
\end{definition}

\noindent\textbf{\textsf{Augmented Sketch (ASketch)~}}
ASketch~\cite{royAugmentedSketchFaster2016} is a CMS extension for skewed streams, where a small subset of keys account for the majority of updates.
It uses a \emph{filter}, \augFilterName{}, to count occurrences of the heaviest keys separately from the \cmsName{} matrix, improving throughput and accuracy.
Alg.~\ref{algo:asketch:insert} shows its operations;
\cmsInsertAndPQFn (\autoref{algo:asketch:insert:cms-update-and-pq}) is a simple extension of the CMS update to also get a point query estimate for the key.
\augFilterName{} can be efficiently searched for a key (\autoref{algo:asketch:insert:simd-scan}), e.g., using SIMD; if the key is found, the update increments just a single filter counter (\autoref{algo:asketch:insert:inc-item-in-filter}), avoiding collisions with light keys and
calculation of \sketchHeight{} hash functions and
increments.
Keys are swapped between filter and CMS matrix as needed, to keep the most frequent keys in the filter (\autoref{algo:asketch:insert:swap}).
Note that \augFilterDrift{} (\autoref{algo:asketch:insert:delta}) is the \emph{exact number of occurrences of \dataKey{} while resident in the filter}.\looseness=-1

\input{algos/asketch-insert-and-query.tex}

%% file: algos/asketch-insert-and-query.tex
\begin{algorithm}
    \caption{%
        ASketch{} -- with highlighted enhancements for \framework{}, detailed in~§~\ref{sec:overview}.%
        \algoHighlightSingleLine{tm:asketch:insert:0}{green}{48pt}{-55.5pt}% Offset to highlight Enhanced only
        \algoHighlightSingleLine{tm:asketch:insert:1}{green}{-1pt}{239pt}% Offset to include side comment
        \algoHighlightSingleLine{tm:asketch:insert:2}{green}{-1pt}{3pt}%
        \algoHighlightSingleLine{tm:asketch:insert:3}{green}{-1pt}{3pt}%
        \algoHighlightSingleLine{tm:asketch:insert-enhanced:update-and-pq-enhanced}{green}{-0.5pt}{0.5pt}%
        \algoHighlightSingleLine{tm:asketch:insert-enhanced:update-enhanced}{green}{-0.5pt}{0.5pt}%
    }
    \label{algo:asketch:insert}

    \Fn{\tikzmark{tm:asketch:insert:0:start}\asketchInsertEnhancedFn{key \dataKey{}, value \dataValue{}}\tikzmark{tm:asketch:insert:0:end}}{\label{algo:asketch:insert:fn}
        \uIf(\Comment*[f]{Increment count in filter}){\dataKey{} in \augFilterName{}}{  \label{algo:asketch:insert:simd-scan}
            \augFilterCount{\dataKey{}} \pluseq{} \dataValue{}\;  \label{algo:asketch:insert:inc-item-in-filter}

            \tikzmark{tm:asketch:insert:1:start}%
            \augFilterProjection{\dataKey{}} $\gets w \dataValue + (1\!-\!w)\augFilterProjection{\dataKey}$%
            \tikzmark{tm:asketch:insert:1:end}%
            \Comment*[r]{$w = 0.8$, bias towards recent changes in skew}  \label{algo:asketch:insert-enhanced:update-ewma}
        }
        \uElseIf(\Comment*[f]{Add key to filter}){\augFilterName{} not full}{
            \augFilterCount{\dataKey{}} $\gets$ \dataValue{}\;
            \augFilterOldCount{\dataKey{}} $\gets$ 0\;

            \tikzmark{tm:asketch:insert:2:start}%
            \augFilterProjection{\dataKey{}} $\gets \dataValue$%
            \tikzmark{tm:asketch:insert:2:end}\;
        }
        \Else(\Comment*[f]{Update sketch}){
            estimatedFreq $\gets$ \cmsName.\cmsInsertAndPQEnhFn{\dataKey{}, \dataValue{}}\;  \label{algo:asketch:insert:cms-update-and-pq}
            \If{$\text{estimatedFreq} > \min_{\text{key in \augFilterName}}(\augFilterCount{\text{key}})$}{  \label{algo:asketch:insert:swap}
                minKey $\gets \argmin_{\text{key in \augFilterName}}(\augFilterCount{\text{key}})$\;
                $\augFilterDrift \gets \augFilterCount{\text{minKey}} - \augFilterOldCount{\text{minKey}}$\;  \label{algo:asketch:insert:delta}
                \cmsName.\cmsInsertEnhFn{minKey, \augFilterDrift}\;  \label{algo:asketch:insert:race-condition-1}
                \augFilterCount{\dataKey{}} $\gets$ estimatedFreq\;  \label{algo:asketch:insert:race-condition-2}
                \augFilterOldCount{\dataKey{}} $\gets$ estimatedFreq\;

                \tikzmark{tm:asketch:insert:3:start}%
                \augFilterProjection{\dataKey{}} $\gets \dataValue$%
                \tikzmark{tm:asketch:insert:3:end}\;
            }
        }
    }

    \Fn{\asketchQueryFn{key \dataKey{}}}{\label{algo:asketch:query:fn}
        \lIf{\dataKey{} in \augFilterName}{%
            \label{algo:asketch:query:heavy-key}
            \Return \augFilterCount{\dataKey{}}%
        }
        \lElse{\Return \cmsName.\pointQueryFn{\dataKey{}}}
    }
\end{algorithm}

%% file: include/03_problem.tex
\section{Problem Description and Analysis}  \label{sec:problem}

We target to estimate item frequencies and frequency moment queries of high-rate data streams
in one pass using a single sketch.
Concurrent operations are \emph{updates} (to process input tuples) and \emph{queries} of various types.
The input is a stream of tuples \dataTuple{}, representing \dataValue{} occurrences of key \dataKey{}.
We consider a \emph{cash register} model, i.e. $\dataValue{} > 0$.
We target multi-core, multi-threaded shared memory systems supporting atomic primitives including fetch-and-add and compare-and-swap.
Threads communicate via a coherent memory model, and do not fail or crash.
A solution needs to balance a multi-way trade-off that we analyze in the following.

\noindent\textbf{\textsf{Accuracy and Consistency~}}
Strong consistency requires atomic views of the data structure state but may involve significant overhead; weakening these semantics \emph{arbitrarily} may reduce accuracy in unclear ways.
We aim for a balance: fresh query results, avoiding delays from unnecessary overhead, and with clear semantics, avoiding the accuracy pitfalls of both.
To facilitate this, we build on the idea of \textit{Intermediate Value Linearizability} (IVL)~\cite{rinbergIntermediateValueLinearizability2023}, that extends \emph{weak regularity}~\cite{nikolakopoulosConsistencyFrameworkIteration2015}, requiring that a concurrent query return a value in the interval between the minimum and maximum ones it could return in any linearization of the execution.
In the cash register model, the considered query values increase monotonically with updates; hence:\looseness=-1
\begin{observation}  \label{obs:monotonic-ivl-is-bounded-between-start-and-end}
    The return value of an IVL query \qop{} of a monotonically increasing function is bounded between the return value of an ideal sequential query at the beginning of \qop{} (only observing completed updates, no overlapping ones) and that of a similar query at the end of~\qop{}.\looseness=-1
\end{observation}

$Q$ must observe \emph{exactly once} all updates completed before it started;
the range of permitted return values, effectively the `inaccuracy' accepted as a consequence of IVL, is due to updates concurrent with $Q$.
As we target high data rates, a \emph{short query latency} is critical for minimizing this range and preserving \emph{freshness}.
To enable better performance, consistency can be further relaxed where tolerable:
\cite{rinbergIntermediateValueLinearizability2023} combines $r$\nobreakdashes-relaxation~\cite{henzingerQuantitativeRelaxationConcurrent2013} with IVL,
-- e.g., allowing \qop{} to also miss up to $r$ updates preceding it --
paraphrased here:

\begin{definition}
\label{defn:r-relaxed-ivlness}
    A query $Q$
    is $r$\nobreakdashes-relaxed IVL if it returns a value between the min and max values it could return in any linearization that may reorder Q with up to $r$ operations.
\end{definition}

\noindent\textbf{\textsf{Multiple Query Types~}}  %
The targeted queries include \emph{local} and \emph{global} ones.
The former involve part of the data, e.g., a point query for a certain key; the latter regard complete contents of the sketch.
Answering such queries associates with bulk operations
\cite{dworkTimeLapseSnapshots1999, nikolakopoulosConsistencyFrameworkIteration2015, petrankLockFreeDataStructureIterators2013}.
Further, supporting multiple queries also concerns their relative consistency;
to reason about associated properties, we identify \emph{operation monotonicity} as a useful notion
to tell us
how aligned the views of different operations are; paraphrasing from~\cite{dworkTimeLapseSnapshots1999, nikolakopoulosConsistencyFrameworkIteration2015}:\looseness=-1

\begin{definition}[Monotonicity of scans]  \label{defn:monotonicity-scans}
    For queries $Q_1$ and $Q_2$, where $Q_1$ precedes $Q_2$ (denoted \precedes{}), all updates observed (i.e., accounted for) by $Q_1$ must also be observed by $Q_2$.\looseness=-1
\end{definition}

\noindent\textbf{\textsf{Accuracy and Memory~}}  %
The \epsdelta{} bounds of sketches allow trading more memory for improved accuracy;
e.g., in CMS, a query can overestimate due to hash collisions --
more counters (or filters as in ASketch) reduce collisions.
In the sequential setting, there are well-studied bounds, which ideally should be preserved under concurrency.
Similarly important are memory \emph{layout and management}.
Partitioning the input domain can aid parallelism as threads can safely summarize each partition concurrently, enhancing accuracy and efficiency, shown for local queries in~\cite{stylianopoulosDelegationSketchParallel2020};
however, challenges arise for global queries which scan multiple partitions.\looseness=-1

\noindent\textbf{\textsf{Summary: Goals and Challenges~}}
\textbf{\textsf{(G1)}} \emph{IVL and associated relaxations} are identified as consistency targets, due to explainability and advantages in synchronization and efficiency.
\emph{Low query latency} becomes a significant goal, allying with \emph{freshness}.
\textbf{\textsf{(G2)}} \emph{Monotonicity of scans} is desirable for multiple query types.
\textbf{\textsf{(G3)}}
\emph{Maintaining a single data structure} for answering multiple queries may enable better use of the memory budget if the accuracy for each query can benefit from the full memory.
We identify \emph{partitioning} as an ancillary approach to improving accuracy and memory utilization, also aiding parallelization.

%% file: include/04_overview.tex
\section{\framework{} -- Design and Coordination}  \label{sec:overview}

We describe the design space and coordination in \framework{} to enable concurrent estimation of multiple queries from a single, composite data structure (Fig.~\ref{fig:overview:data-structure}) in conjunction with partitioning, as motivated in §~\ref{sec:problem}, for consistency, memory efficiency, and accuracy-friendliness.

\begin{figure}[!htb]
    \includeinkscape[pretex=\small, width=\linewidth]{overview_fig_v3.pdf_tex}
    \caption{
        \framework{}'s composite data structure.
        Partitions, horizontal slices of memory, are modifiable only by the owner thread.
        Delegation filters $DF$, on the right, are periodically handed over to the owning thread, $\delFilterSlots{} = 16$.
        All updates eventually reach partition-local ASketches (\sketchName{} and \augFilterName{}), to the left of the vertical boundary.
        Dotted components are for optimizations.%
    }
    \label{fig:overview:data-structure}
\end{figure}

\begin{table}[!htb]
    \caption{Notation used throughout.}\label{tab:notation}
    \begin{minipage}[t]{.5\linewidth}
        \vspace{0pt}
        \begin{tabular}{c @{\hskip\tabcolsep} p{4.5cm}}
            \toprule
                                  \thead{Symbol}                       & \thead{Description}                                          \\
            \midrule
                                  \delThreads{}                        & No. of partitions \& threads                      \\
                         \sketchWidth{}, \sketchHeight{}               & CMS rows and columns                 \\
                                \delFilterSlots{}                      & No. of slots in ASketch and delegation filters            \\
                               \delFilterMaxCount{}                    & Max. no. of updates buffered in delegation filter         \\
            \midrule
                              \delOwner{\dataKey{}}                    & Get partition owning key \dataKey{}                          \\
                          \delThreadI{\delThreadIIndex}                & Updater thread for partition \delThreadIIndex{}  \\
                        \delThreadSketch{\delThreadIIndex}             & Local sketch for \delThreadI{\delThreadIIndex}                \\
            \delThreadI{\delThreadIIndex}.\delThreadInsertedElements{} & No. of updates completed by \delThreadI{\delThreadIIndex} \\
            \bottomrule
        \end{tabular}
    \end{minipage}\hfill
    \begin{minipage}[t]{.5\linewidth}
        \vspace{0pt}
        \begin{tabular}{c @{\hskip\tabcolsep} p{4.5cm}}
            \toprule
                                 \thead{Symbol}                      & \thead{Description}                                                                                         \\
            \midrule
                        \sketchName{}.\cmsPartialFtwo{}              & Sum for \cmp{} per row (Def.~\ref{defn:cm+})                                    \\
                     \delThreadAugFilter{\delThreadIIndex}           & ASketch filter for \delThreadI{\delThreadIIndex}                                                            \\
            \makecell[t]{\augFilterCount{\dataKey},\\%
                         \augFilterOldCount{\dataKey}}               & Count \& Old count of heavy key \dataKey{} in \augFilterName{}                                      \\
                         \augFilterProjection{\dataKey}              & Projected count of buffered occurrences of heavy key \dataKey{}                                                       \\
            \delThreadDelFilter{\delThreadIIndex}{\delThreadIIIndex} & Delegation filter of \delThreadI{\delThreadIIndex} for updates owned by partition \delThreadIIIndex{} \\
                          \delFilterCount{\delThreadIIIndex}{\dataKey}                & Count of key \dataKey{} buffered in \delFilterTarget{\delThreadIIIndex{}}                             \\
            \makecell[t]{\delFilterTarget{\delThreadIIIndex{}}.\delFilterFilledSlots{},\\[-1pt]%
                  \delFilterTarget{\delThreadIIIndex{}}.\delFilterBufferedCount{}}        & Number of keys and updates buffered in \delFilterTarget{\delThreadIIIndex{}}                                                     \\
            \bottomrule
        \end{tabular}
    \end{minipage}
\end{table}

The input domain is split into \delThreads{} partitions, as is the data structure (horizontal slices in Fig.~\ref{fig:overview:data-structure}).
Updates are performed by \delThreads{} threads, each `owning' one partition (ASketch and other components), with update access to it.
When thread \delThreadI{\delThreadIIndex} processes an update for a key owned by another partition, the update is delegated (i.e., the opposite of work-stealing) to the owning thread (Alg.~\ref{algo:delsketch:insert}).
\emph{Delegation filters} buffer updates to reduce synchronization overhead.
Filters behave as maps of \delFilterSlots{} keys, small enough to be searched efficiently via SIMD operations.
Periodically (upon a triggering condition, \autoref{algo:delsketch:insert:if-filter-full-handover}) filters are pushed to the owner to be flushed: contents are transferred to its local ASketch in bulk (Alg.~\ref{algo:delsketch:ppi}).

The domain-partitioning and delegation idea is also the basis of the \delsketch{}~\cite{stylianopoulosDelegationSketchParallel2020} for point queries,
summarized in §~\ref{sec:overview:pq} for self-containment.
Moreover, we here show their consistency properties as a query type in \framework{}.
However, the local approach of assigning the query to the owner thread will not work for global queries which span all partitions.
We investigate the trade-offs relating concurrency and accuracy for such bulk queries, as identified in §~\ref{sec:problem}:
the extremes of the spectrum serve as baselines,
followed by our balancing approach.\looseness=-1

\begin{figure}[!hbt]
\vspace{.05\baselineskip}
\begin{minipage}[t]{.47\linewidth}
    \input{algos/delsketch-insert.tex}
\end{minipage}\hfill%
\begin{minipage}[t]{.485\linewidth}
    \input{algos/delsketch-ppi.tex}
\end{minipage}
\end{figure}

\underline{\fullsync{}}:
A baseline for strong consistency, using a Readers-Writer lock to allow shared access for updates and point queries, but exclusive access for global queries (acting as writers) to see a \emph{consistent global state}.
\fone{} is estimated by summing all counters on any row of each \delThreadSketch{\delThreadIIndex} and counts in all filters, yielding the number of updates completed before the query.
\ftwo{} can be estimated using \cmp{} by merging all sketches and filter contents, thus preserving its error bounds.
Note several drawbacks: updates incur locking overhead even when no global query takes place; further,
while thread-safe, the result can be stale due to `stopping the world' -- input tuples keep arriving but are not processed or visible to the query.\looseness=-1

\underline{\nosync{}} (§~\ref{sec:overview:elem:nosync}):
This approach targets optimistic synchronization, for exploring freshness and concurrency maximization, at the cost of consistency, possibly risking mis-calculations.

\underline{\lagom{}} (§~\ref{sec:overview:lagom}): Our balanced method for concurrent queries with low latency, `just-enough' calculation, and lightweight synchronization.
Query results have stronger semantics than \nosync{}, and are based on fresher state than \fullsync{}.

\subsection{Point Queries}  \label{sec:overview:pq}
Point queries in the partitioned design (Alg.~\ref{algo:delsketch:pq}) follow~\cite{stylianopoulosDelegationSketchParallel2020}.
A point query for key \dataKey{} is answered by \delThreadI{\delOwner{\dataKey{}}}, the thread owning \dataKey{}, estimating \frequencyEst{\dataKey{}} as the sum of occurrences of \dataKey{} in the thread-local ASketch of \delThreadI{\delOwner{\dataKey{}}} (\autoref{algo:delsketch:pq:query-sketch}) and in relevant delegation filters of other threads (\autoref{algo:delsketch:pq:query-dfs}).
Skew in the input data is beneficial to \frequencyEst{\dataKey{}}'s accuracy;
frequent keys are often present in the filters where they are counted accurately, akin to ASketch.
Based on the IVL definition and Obs.~\ref{obs:monotonic-ivl-is-bounded-between-start-and-end}, along the fact that entries in the partition-local sketch are increasing with subsequent updates, and that the thread owning a key handles the realization of its updates and queries (hence cannot miss completed updates or double-count any), we have:\looseness=-1

\begin{restatable}{lemma}{lemmaPQisIVL}
    \label{lemma:pq-is-ivl}%
    \delsketch{}-based \pq{} is an IVL implementation of ASketch point query.\looseness=-1
\end{restatable}

\subsection{\nosync{} for Bulk Queries}  \label{sec:overview:elem:nosync}
We explore a baseline with no synchronization between updates and global queries, to explore freshness and  concurrency maximization (at the cost of consistency).

\noindent\textbf{\textsf{\textfone{}~}}
To estimate \fone{}, improving efficiency and freshness compared to an approach as in \fullsync{},
we introduce partial results in form of per-thread counters (\mbox{\delThreadI{\delThreadIIndex}.\delThreadInsertedElements{}}) for the number of performed updates (\autoref{algo:lagom:insert:inc-numinserts}).
The \fonenosync{} query (Eq.~\ref{eqn:nosync-f1}) sums these counters, improving locality, efficiency, and NUMA-friendliness compared to \fullsync{}.
\begin{equation}  \label{eqn:nosync-f1}
    \fonenosync = \sum_{\delThreadIIndex}^{\delThreads} \delThreadI{\delThreadIIndex}.\text{\delThreadInsertedElements}
\end{equation}

\begin{restatable}{lemma}{lemmaNosyncFOneIsIVL}
    \label{lemma:nosync-f1-is-ivl}%
    \fonenosync{} estimates \fone{} with IVL semantics.
\end{restatable}

\begin{figure}[htb!]
\noindent
\begin{minipage}[t]{.48\linewidth}
    \input{algos/delsketch-pq.tex}
    \input{algos/nosync-f2.tex}
\end{minipage}\hfill%
\begin{minipage}[t]{.47\linewidth}
    \vspace*{-3pt}
    \input{algos/lagom-f2.tex}
\end{minipage}
\end{figure}

\noindent\textbf{\textsf{\textftwo{}~}}
A global \ftwoest{} equals the sum of per-partition \ftwoest{}, as partitions contain independent, non-overlapping parts of the input.
However, an approach as for \textfone{} with per-partition partial results poses obstacles:
threads would update the partial result at the \emph{owning partition},
reintroducing contention and serialization on shared data into the delegation design.

Instead, \analysisOursFullSeq{} (Eq.~\ref{eqn:nosync-f2}) uses \cmp{} to estimate \ftwo{} for each \delThreadSketch{\delThreadIIndex}.
Then, for each key~\dataKey{} in \delThreadAugFilter{\delThreadIIndex}, \heavyKeyFTwoNosync{} (Eq.~\ref{eqn:nosync-f2-heavy-keys}) performs a point query (Alg.~\ref{algo:delsketch:pq}, using the fast path on \autoref{algo:asketch:query:heavy-key}) to obtain $\frequencyEst{\dataKey{}}^2$, the \ftwo{} contribution of \dataKey{}.
However, some occurrences of \dataKey{} may be stored in \delThreadSketch{\delThreadIIndex} and their contribution included in the \cmp{} estimate; this needs to be subtracted.
We provide a geometric intuition for this calculation in §~\ref{sec:analysis:delegation} and Fig.~\ref{fig:squares:delegation}.
Alg.~\ref{algo:nosync:f2} shows the synchronization (or, rather, lack thereof) for concurrent \analysisOursFullSeq{}.

\begin{align}
    \analysisOursFullSeq &=
    \sum_{\delThreadIIndex}^{\delThreads} \left(
        \ftwoestm{\cmp}\!(\delThreadSketch{\delThreadIIndex})
        + \sum_{\text{\dataKey{}}}^{\delThreadAugFilter{\delThreadIIndex}}
            \heavyKeyFTwoNosync(\dataKey{})
    \right) \label{eqn:nosync-f2}\\
    \heavyKeyFTwoNosync(\dataKey) &= \left(
        \delThreadAugFilterCount{o}{\dataKey}
        + \sum_{\delThreadIIIndex}^{\delThreads}
            \delThreadDelFilterCount{\delThreadIIIndex}{o}{\dataKey}
    \right)^{2}
    \!-\left(\delThreadAugFilterOldCount{o}{\dataKey}\right)^2
    && \text{where}\ o = \delOwner{\dataKey{}}
    \label{eqn:nosync-f2-heavy-keys}
\end{align}

\begin{restatable}{observation}{obsNosyncFTwoIsUnsafe}
    \label{obs:nosync-f2-is-unsafe}%
   An \ftwoestm{\nosync} query $Q$ can miss or double-count updates, due to data movement by overlapping  processing of delegated updates (Alg.~\ref{algo:lagom:ppi}).
\end{restatable}

While \nosync{} avoids the overhead of \fullsync{}, serving as a baseline for maximal concurrency and exploration of freshness, the lack of synchronization leaves weak consistency guarantees for \ftwo{} estimations\footnotemark{}, implying arbitrary fluctuations from the accuracy bounds.
\footnotetext{To be precise, \ftwoestm{\nosync} is quiescence-consistent~\cite{herlihyArtMultiprocessorProgramming2021}: in absence of concurrent updates, the `bad things' in \autoref{obs:nosync-f2-is-unsafe} cannot happen. But this property is to little purpose in high-rate scenarios.}

\subsection{\lagom{} for Bulk Queries -- \textftwo{}}  \label{sec:overview:lagom}
We now present our design to enable high throughput with clear concurrency semantics.
We build upon some of the described components, i.e. support updates via Alg.~\ref{algo:delsketch:insert}, with highlighted enhancements, point queries via Alg.~\ref{algo:delsketch:pq} (§~\ref{sec:overview:pq}) and \foneest{} via Eq.~\ref{eqn:nosync-f1} (§~\ref{sec:overview:elem:nosync}).
The crux to efficient, concurrent \ftwoest{} queries with consistency and accuracy guarantees lies in two key ideas:
(1)~\emph{efficient maintenance of partial results}, which, based on a geometric observation, allows to argue about accuracy relative to the sequential bounds; and
(2)~\emph{lightweight synchronization} implying IVL, over only few variables to scan, due to how partial results are maintained.

\noindent\textbf{\textsf{Partial results for \textftwo{}~}}
While a global \ftwoest{} can be obtained from summing per-partition \ftwoest{} values as in \ftwoestm{\nosync}, maintaining these per partition is, unlike \fone{}, not straightforward.
Instead, we compute partial results to simplify the heaviest operations in Eq.~\ref{eqn:nosync-f2}, \ref{eqn:nosync-f2-heavy-keys}, which are:
(1)~\cmp{} on the underlying CMS, that requires reading all $\sketchHeight\!\times\!\sketchWidth$ counters of the sketch, and
(2)~\heavyKeyFTwoNosync{} (Eq.~\ref{eqn:nosync-f2-heavy-keys}), that reads all delegation filters for each heavy key.

For (1), we maintain the per-row sums in Def.~\ref{defn:cm+} for each sketch row, labeled \cmsPartialFtwo{} in Fig.~\ref{fig:overview:data-structure}.
This is done within the enhanced CMS update in Alg.~\ref{algo:asketch:insert}, with incremental (associative) calculations.
\ftwoestm{\cmp} can then return the $\min$ of these \sketchHeight{} values (highlighted in Eq.~\ref{eqn:lagom-f2}).

For (2), we aim to avoid scanning delegation filters when estimating the frequency of heavy keys.
However, as filters can buffer a significant number of updates, particularly for skewed streams, their contents should not be ignored, to avoid excessive underestimation (analyzed in §~\ref{sec:analysis:delegation}).
To compensate, we devise \heavyKeyFTwoLagom{} (Eq.~\ref{eqn:lagom:f2-heavy-keys}) as a lightweight  estimation of the number of heavy-key occurrences buffered in delegation filters.
For each slot, ASketch filters maintain an exponentially weighted moving average of the number of occurrences received during filter flushes (\autoref{algo:asketch:insert-enhanced:update-ewma}, invoked from \autoref{algo:lagom:ppi:asketch-insert-enhanced}).
We replace the scan over delegation filters in \heavyKeyFTwoNosync{} with the following idea (highlighted in Eq.~\ref{eqn:lagom:f2-heavy-keys}), requiring only one read from the ASketch filter:
there are \delThreads{} delegation filters for a partition, projected to contain \augFilterProjection{\dataKey{}} counts of \dataKey{} once full and getting flushed.
At any point when the query scans a partition, filters are on average half full.
This yields \analysisOursProjSeq{} (Eq.~\ref{eqn:lagom:f2-heavy-keys}).
In §~\ref{sec:analysis} we analyze the accuracy implications and the association with the sequential bounds.

\begin{align}
    \analysisOursProjSeq &=
    \sum_{\delThreadIIndex}^{\delThreads} \left(%
        \tikzmark{tm:lagom:f2-cmplus-opt:start}%
        \min(\delThreadSketch{\delThreadIIndex}.\text{\cmsPartialFtwo})%
        \tikzmark{tm:lagom:f2-cmplus-opt:end}
        + \sum_{\text{\dataKey{}}}^{\delThreadAugFilter{\delThreadIIndex}}
            \heavyKeyFTwoLagom(\dataKey{})
    \right) \label{eqn:lagom-f2}\\
    \heavyKeyFTwoLagom(\dataKey) &= \left(
        \delThreadAugFilterCount{o}{\dataKey}
        +\tikzmark{tm:lagom:f2-heavy-keys:start}\delThreads\!\cdot\!\frac{ \delThreadAugFilterProjection{o}{\dataKey}}{2}\tikzmark{tm:lagom:f2-heavy-keys:end}
    \right)^{2}
    \!-\left(\delThreadAugFilterOldCount{o}{\dataKey}\right)^2
    && \text{where}\ o = \delOwner{\dataKey{}} \label{eqn:lagom:f2-heavy-keys}
\end{align}
\algoHighlight{tm:lagom:f2-cmplus-opt}{green}{{0pt,10pt}}{{1pt,-4pt}}
\algoHighlight{tm:lagom:f2-heavy-keys}{green}{{-1pt,17pt}}{{0pt,-10pt}}

\noindent\textbf{\textsf{Lightweight Synchronization~}}
The \ftwoestm{\lagom} query (Alg.~\ref{algo:lagom:f2}) uses lightweight synchronization to safely calculate \analysisOursProjSeq{} concurrently with updates.
To achieve the IVL-semantics goal (G1) set in §~\ref{sec:problem} and avoid the safety problems of \ftwoestm{\nosync} seen in Obs.~\ref{obs:nosync-f2-is-unsafe}, \ftwoestm{\lagom} synchronizes with each partition \delThreadIIndex{} being scanned (i.e., with thread \delThreadI{\delThreadIIndex}) via a handshake, involving a pair of initially equal version numbers (as in~\cite{lamportConcurrentReadingWriting1977}) for detecting a concurrent filter flush, and an atomic flag set by the query, to signal when the partition is being scanned.

In the common case, when \delThreadI{\delThreadIIndex} processes a tuple \dataTuple{}, Alg.~\ref{algo:delsketch:insert} with enhancements finds available space in \delThreadDelFilter{\delThreadIIndex}{\delOwner{\dataKey}} (condition on \autoref{algo:delsketch:insert:while-filter-still-full});
\delThreadDelFilterCount{\delThreadIIndex}{\delOwner{\dataKey}}{\dataKey} is then incremented by \dataValue{} (\autoref{algo:delsketch:insert:inc-item-in-filter} or \ref{algo:delsketch:insert:add-item-to-filter}), but a handover is \emph{not} triggered immediately (\autoref{algo:delsketch:insert:if-filter-full-handover} is false).
\ftwoestm{\lagom} queries cannot directly account for this update before \delThreadDelFilter{\delThreadIIndex}{\delOwner{\dataKey}} is flushed to \delThreadI{\delOwner{\dataKey}}.
\ftwoestm{\lagom} targets a consistent view of each partition's relevant information independently.
Concurrent filter flushes (Alg.~\ref{algo:lagom:ppi}) are detected if version numbers mismatch (\autoref{algo:lagom:f2:versions-match}), causing the query to retry, though only for at most one concurrent flush per partition, as subsequent flushes will be stalled by the flag (\autoref{algo:lagom:ppi:flag-check}).
The sum of the collected per-partition partial results is then returned as \ftwoest{}.
The update-query interaction implies:

\begin{restatable}{lemma}{lemmaLagomAtomicSnapshotAndNoDeadlocks}
    \label{lemma:lagom-is-good}%
    \ftwoestm{\lagom} gets an atomic snapshot per partition, and cannot deadlock with updates.\looseness=-1
\end{restatable}

\noindent\textbf{\textsf{Optimizations on Filters: Bounds and Self-Delegation~}}
To bound the interval between filter flushes, a parameter \delFilterMaxCount{} limits the buffering capacity of filters (\autoref{algo:lagom:insert:inc-item-count-existing} and \ref{algo:lagom:insert:inc-item-count-new}).
Lower values for \delFilterMaxCount{} trade performance for accuracy -- both for updates, as filters are unavailable more often, and for queries, where more overlapping flushes interfere with them.
The impact of \delFilterMaxCount{} is studied in §~\ref{sec:analysis:delegation},
according to this observation, motivating tracking of updates in \ftwoestm{\lagom}:\looseness=-1
\begin{observation}  \label{obs:delegation-filters-buffer-r-updates}
    At most $r = \delThreads \delFilterMaxCount$ single increment updates can be buffered in delegation filters and may thus not be explicitly observed by \ftwoestm{\lagom} queries.
\end{observation}

Further, in \delsketch{}, a thread processing an update for a key it owns updates its ASketch directly.
However, in \lagom{}, ASketch updates require synchronizing with global queries, causing delays particularly in partitions owning heavy keys.
Instead, each partition is extended with a \emph{self-delegation filter} to buffer local updates,
targeting efficiency improvements, as updates are aggregated and flushed in bulk, and also interfere less with queries.\looseness=-1

\subsection{Multiple Query Types -- Relative Consistency}  \label{sec:overview:query-consistency-monotonicity}
In §~\ref{sec:problem} we identify \emph{monotonicity of scans}~\cite{dworkTimeLapseSnapshots1999} (G2) as intuitive and desirable behavior -- a query should observe the updates observed by a previously performed query.
Having a single data structure implies certain monotonicity compared to a disjoint per-sketch data structure.
As \framework{} consists of multiple components, an update can become visible to queries of different types at slightly different points.
Here we bound how much/little `the world can differ' between query types.
To this end, we identify as sub-operations of an update $U$ the atomic operations, denoted here \subop{U}{op}, after which $U$ becomes visible to a query of the respective type:
\subop{U}{\pq} at \autoref{algo:delsketch:insert:inc-item-in-filter} or \ref{algo:delsketch:insert:add-item-to-filter},
\subop{U}{\fone{}} at \autoref{algo:lagom:insert:inc-numinserts}, and
\subop{U}{\ftwo} at \autoref{algo:lagom:ppi:ts2}.
We have $\subop{U}{\pq} \precedes \subop{U}{\fone} \precedes \subop{U}{\ftwo}$ from program order.
The cross-query consistency properties are:

\begin{restatable}{lemma}{lemmaMonotonicityOfLagomQueries}
    \label{lemma:monotonicity-of-queries}%
    For queries $Q_1 \precedes Q_2$
    the monotonicity of scans for each combination follows \looseness=-1

    \centering
    \begin{tabular}{ccccc}
        \toprule
         & \theadfont{Q\textsubscript{2} \textrightarrow} &   \pq{}   &           \textfone{}           &           \textftwo{}            \\\addlinespace[0.3\defaultaddspace]
        &       \pq{}       & Monotonic & $\delThreads$-relaxed monotonic &      $r$-relaxed monotonic       \\
        \theadfont{Q\textsubscript{1}} &    \textfone{}    & Monotonic &            Monotonic            & $r\delThreads$-relaxed monotonic \\
        &    \textftwo{}    & Monotonic &            Monotonic            &            Monotonic             \\
        \bottomrule
    \end{tabular}
\end{restatable}

Note that:
(1)~The bounds are pessimistic, e.g. for \textpq{} \precedes{} \textfone{} to deviate by \delThreads{} updates, all must occur in the partition queried by $Q_1$, while for all other updates except these, i.e., where $\subop{U}{\fone{}} \precedes Q_2$, full monotonicity of scans applies.
(2)~\delThreads{} is small relative to update rates.
(3)~Due to the compensation scheme of \ftwoestm{\lagom}, even in the very unlikely case that the maximum deviation in operations observed occurs, actual results differ significantly less.\looseness=-1

%% file: algos/delsketch-insert.tex
\begin{algorithm}[H]
    \caption{\delsketch{} update on~\delThreadI{\delThreadIIndex}. Enhancements for \lagom{} in green.\looseness=-1}
    \label{algo:delsketch:insert}
    \Fn{\insertAlgoFn{key \dataKey{}, value \dataValue{}}}{
%        owner $\gets$ \findOwnerFn{\dataKey{}}\;
        filter $\gets$ \delThreadDelFilter{\delThreadIIndex}{\delOwner{\dataKey{}}}\;

        \While{filter.\delFilterFilledSlots{} = \delFilterSlots{} \tikzmark{tm:lagom:insert:1:start}or filter.\delFilterBufferedCount{} $\mathit{\geq}\ \delFilterMaxCount$\tikzmark{tm:lagom:insert:1:end}}{  \label{algo:delsketch:insert:while-filter-still-full}
            \hyperref[algo:delsketch:ppi:fn]{\processPendingInsertsFn{}}\;
        }

%        \eIf(\Comment*[f]{Increment count in filter}){\dataKey{} in filter}{
        \eIf{\dataKey{} in filter}{
            filter[\dataKey{}] \pluseq{} \dataValue{}\;  \label{algo:delsketch:insert:inc-item-in-filter}
            \tikzmark{tm:lagom:insert:2:start}%
            filter.\delFilterBufferedCount{} \pluseq{} \dataValue{}%
            \tikzmark{tm:lagom:insert:2:end}\;  \label{algo:lagom:insert:inc-item-count-existing}
%        }(\Comment*[f]{Add key to filter}){
        }{
            filter[\dataKey{}] $\gets \dataValue$\;  \label{algo:delsketch:insert:add-item-to-filter}
            filter.\delFilterFilledSlots{} \pluseq{} 1\;
            \tikzmark{tm:lagom:insert:3:start}%
            filter.\delFilterBufferedCount{} \pluseq{} \dataValue{}%
            \tikzmark{tm:lagom:insert:3:end}\;  \label{algo:lagom:insert:inc-item-count-new}
        }

%        \delThreadI{\delThreadIIndex}.\delThreadInsertedElements{}\plusplus{}%
        \tikzmark{tm:lagom:insert:4:start}%
        \delThreadI{\delThreadIIndex}.\delThreadInsertedElements{} \pluseq{} \dataValue{}%
        \tikzmark{tm:lagom:insert:4:end}\;  \label{algo:lagom:insert:inc-numinserts}

        \If{filter.\delFilterFilledSlots{} = \delFilterSlots{} \tikzmark{tm:lagom:insert:5:start}or filter.\delFilterBufferedCount{} $\mathit{\geq}\ \delFilterMaxCount$\tikzmark{tm:lagom:insert:5:end}}{  \label{algo:delsketch:insert:if-filter-full-handover}
            owner.\delThreadFullFilters{}.push(filter)\;
        }
    }
\end{algorithm}
% Apply highlights
\algoHighlightSingleLine{tm:lagom:insert:1}{green}{-1pt}{1pt}
\algoHighlightSingleLine{tm:lagom:insert:2}{green}{-1pt}{3pt}
\algoHighlightSingleLine{tm:lagom:insert:3}{green}{-1pt}{3pt}
%\algoHighlightSingleLine{tm:lagom:insert:4}{green}{-1pt}{3pt}
\algoHighlight{tm:lagom:insert:4}{green}{{-1pt,7.5pt}}{{3pt,-2pt}}
\algoHighlightSingleLine{tm:lagom:insert:5}{green}{-1pt}{1pt}

%% file: algos/delsketch-ppi.tex
\begin{algorithm}[H]
    % Without parbox, caption is too narrow
%    \parbox{\linewidth}{%
        \caption{%
            \parbox[t]{\textwidth}{Processing of delegated updates}
            on~\delThreadI{\delThreadIIndex}. Enhancements for \lagom{} in green.%
        }%
%    }%
    \label{algo:delsketch:ppi}
    \label{algo:lagom:ppi}
    \Fn{\processPendingInsertsFn}{\label{algo:delsketch:ppi:fn}
        \While{\delThreadI{\delThreadIIndex{}}.\delThreadFullFilters{} is not empty}{%

            \tikzmark{tm:lagom:ppi:0:start}%
            Wait until \delThreadI{\delThreadIIndex{}}.\delThreadFlag{} = false%
            \tikzmark{tm:lagom:ppi:0:end}\;  \label{algo:lagom:ppi:flag-check}

            \tikzmark{tm:lagom:ppi:1:start}%
            \delThreadTSone{\delThreadIIndex}\plusplus{}%
            \tikzmark{tm:lagom:ppi:1:end}\;  \label{algo:lagom:ppi:ts1}

            filter $\gets$ \delThreadI{\delThreadIIndex}.\delThreadFullFilters.pop()\;

            \ForEach{\dataKey{} in filter}{%
                \delThreadI{\delThreadIIndex}.\hyperref[algo:asketch:insert:fn]{\asketchInsertEnhancedFn{\dataKey{}, filter[\dataKey{}]}}%
                \label{algo:delsketch:ppi:asketch-insert}%
                \label{algo:lagom:ppi:asketch-insert-enhanced}%
            }

            Clear filter contents\;  \label{algo:delsketch:ppi:clear-filter}
            filter.\delFilterFilledSlots{} $\gets 0$\;

            \tikzmark{tm:lagom:ppi:2:start}%
            filter.\delFilterBufferedCount{} $\gets 0$%
            \tikzmark{tm:lagom:ppi:2:end}\;

            \tikzmark{tm:lagom:ppi:3:start}%
            \delThreadTStwo{\delThreadIIndex}\plusplus{}%
            \tikzmark{tm:lagom:ppi:3:end}\;  \label{algo:lagom:ppi:ts2}
        }
    }
\end{algorithm}

\algoHighlightSingleLine{tm:lagom:ppi:0}{green}{-1pt}{3pt}
\algoHighlight{tm:lagom:ppi:1}{green}{{-1pt,6pt}}{{3pt,-2pt}}
\algoHighlightSingleLine{tm:lagom:ppi:asketch-update-enhanced}{green}{-0.5pt}{0.5pt}
\algoHighlightSingleLine{tm:lagom:ppi:2}{green}{-1pt}{3pt}
\algoHighlight{tm:lagom:ppi:3}{green}{{-1pt,6pt}}{{3pt,-2pt}}

%% file: algos/delsketch-pq.tex
\begin{algorithm}[H]
    \caption{\parbox[t]{\textwidth}{Point query on \delThreadI{o}, $o\!=\!\delOwner{\dataKey{}}$.}}
    \label{algo:delsketch:pq}

    \Fn{\pointQueryFn{key \dataKey{}}}{
        result $\gets$ \delThreadI{o}.\hyperref[algo:asketch:query:fn]{\asketchQueryFn{\dataKey{}}}\;  \label{algo:delsketch:pq:query-sketch}
        \lForEach{\delThreadI{\delThreadIIIndex}}{%
            result \pluseq{} \delThreadDelFilterCount{\delThreadIIIndex}{o}{\dataKey{}}%
            \label{algo:delsketch:pq:query-dfs}%
        }
        \Return result\;
    }
\end{algorithm}

%% file: algos/nosync-f2.tex
\begin{algorithm}[H]
    \caption{\ftwoestm{\nosync} on own thread.}
    \label{algo:nosync:f2}

    \Fn{\queryFTwoNosyncFn}{
        $\mathrm{result} \gets 0$\;

        \ForEach{\delThreadI{\delThreadIIndex}}{  \label{algo:nosync:f2:foreach-partition}

            result $\pluseq{}\ \ftwoestm{\cmp}\!(\delThreadSketch{\delThreadIIndex})$\;     \label{algo:nosync:f2:get-sketch-cmplus}

            \ForEach{\dataKey{} in \delThreadAugFilter{\delThreadIIndex}}{   \label{algo:nosync:f2:for-each-heavy-key}
%                \Comment{Get occurrences of \dataKey{} in \augFilterName{}}
                aFreq $\gets$ \delThreadAugFilterCount{\delThreadIIndex}{\dataKey{}}\;    \label{algo:nosync:f2:get-heavy-key-augfilter}

                \lForEach{\delThreadI{\delThreadIIIndex}}{%
                    \label{algo:nosync:f2:heavy-key-foreach-df}
%                    \Comment{Scan all \delFilterName{} for this partition for buffered occurrences of \dataKey{}}
                    aFreq \pluseq{} \delThreadDelFilterCount{\delThreadIIIndex}{\delThreadIIndex}{\dataKey}%
%                    totalFrequency \pluseq{} \delThreadDelFilter{\delThreadI{\delThreadIIIndex}}{\delThreadIIndex}.query(\dataKey{})\;    \label{algo:nosync:f2:heavy-key-query-df}
                }
                result \pluseq{} $\mathrm{aFreq}^2 - (\delThreadAugFilterOldCount{\delThreadIIndex}{\dataKey})^2$\;
%                $\mathrm{result} \pluseq{} \mathrm{totalFrequency}^2 - \mathrm{\delThreadAugFilter{\delThreadI{\delThreadIIndex}}.oldCount}[\dataKey{}]$\;
            }
        }
        \Return result\;
    }
\end{algorithm}

%% file: algos/lagom-f2.tex
% \DecMargin{1em}
\begin{algorithm}[H]
    \caption{\ftwoestm{\lagom} on own thread.}\label{algo:lagom:f2}

    \Fn{\queryFTwoLagomFn}{
        result $\gets 0$\;

        \ForEach{\delThreadI{\delThreadIIndex}}{  \label{algo:lagom:f2:foreach-partition}
            \delThreadI{\delThreadIIndex{}}.\delThreadFlag{} $\gets$ true\;  \label{algo:lagom:f2:set-flag}

            \Repeat{%
                % End condition
                v1 = v2%
                \label{algo:lagom:f2:versions-match}%
            }{
                % Repeat block
                v2 $\gets \delThreadTStwo{\delThreadIIndex}$\;  \label{algo:lagom:f2:get-v2}
                localResult $\gets \min\!\left(\delThreadSketch{\delThreadIIndex}.\text{\cmsPartialFtwo} \right)$\;     \label{algo:lagom:f2:get-sketch-cmplus}

                \ForEach{\dataKey{} in \delThreadAugFilter{\delThreadIIndex}}{   \label{algo:lagom:f2:for-each-heavy-key}
                    %                \Comment{Get occurrences of \dataKey{} in \augFilterName{}}
                    aFreq $\gets$ \delThreadAugFilterCount{\delThreadIIndex}{\dataKey{}}\;    \label{algo:lagom:f2:get-heavy-key-augfilter}
                    aFreq $\pluseq{}\ \delThreads{} \cdot \delThreadAugFilterProjection{\delThreadIIndex}{\dataKey{}}\ /\ 2$\;    \label{algo:lagom:f2:get-heavy-key-ewma}
                    \mbox{localResult $\smash{\pluseq{}\ \mathrm{aFreq}^2 - (\delThreadAugFilterOldCount{\delThreadIIndex}{\dataKey})^2}$\;}
%                    $\mathrm{partitionResult} \pluseq{} \mathrm{totalFrequency}^2 - \mathrm{\delThreadAugFilter{\delThreadI{\delThreadIIndex}}.oldCount}[\dataKey{}]$\;
                }

                v1 $\gets \delThreadTSone{\delThreadIIndex}$\;  \label{algo:lagom:f2:get-v1}
            }(\Comment*[f]{Retry until match})

            \delThreadI{\delThreadIIndex{}}.\delThreadFlag{} $\gets$ false\;  \label{algo:lagom:f2:clear-flag}

            result \pluseq{} localResult\;  \label{algo:lagom:f2:sum-partition-results}
        }

        \Return result\;
    }
\end{algorithm}
% \IncMargin{1em}

%% file: include/05_analysis.tex
\section{Accuracy of \textftwo{} Estimation}  \label{sec:analysis}

We analyze the consistency and accuracy of \framework{} for \ftwoest{}, as done for \pq{} and \fone{} queries in §~\ref{sec:overview:pq} and §~\ref{sec:overview:elem:nosync}.
We develop a sequence of auxiliary designs (Table~\ref{tab:analysis:names}) and argue for their properties, ultimately arriving at \ftwoestm{\lagom}.
We develop a geometric interpretation of \ftwoest{} in terms of areas of squares, illustrating accuracy properties.
For each step, we describe the organization of the data structure and the \ftwo{} query on it in a sequential setting along with its accuracy, followed by a parallelized construction and its concurrency semantics.

\begin{figure}[!htb]
\begin{minipage}{0.53\textwidth}
    \centering
    \captionsetup{hypcap=false}
    \captionof{table}{Sequential (for accuracy reasoning) \& concurrency-aware \ftwoest{} (for IVL reasoning).}%
    \begin{tabular}{l @{\hspace{3pt}} l @{\hspace{3pt}} l}
        \toprule
        \thead{Data structure}                                         & \thead{Sequential}            & \thead{Concurrent}         \\
        \midrule
        \hyperref[algobox:wide]{\analysisWide{}} (CMS)                 & \cmp{}                        & \analysisWideConc{}        \\
        \hyperref[algobox:partitioned-cms]{\analysisPart{}}            & \analysisPartSeq{}            & \analysisPartConc{}        \\
        \hyperref[algobox:partitioned-asketch]{\analysisPartASketch{}} & \analysisPartASketchSeq{}     & \analysisPartASketchConc{} \\
        \hyperref[algobox:ours]{\analysisOurs{}}                       & \analysisOursFullSeq{}        & \analysisOursFullConc{}    \\
        \hyperref[algobox:ours]{\analysisOurs{}}                       & \analysisOursRelaxedSeq{}(\fwshort{})     & \analysisOursRelaxedConc{} \\
        \hyperref[algobox:ours]{\analysisOurs{}}                       & \analysisOursProjSeq{}        & \analysisOursProjConc{}    \\
        \bottomrule
        \label{tab:analysis:names}
    \end{tabular}
\end{minipage}\hfill%
\begin{minipage}{0.45\textwidth}
    \vspace{.1\baselineskip}
    \captionsetup{type=figure}
    \centering
    \subcaptionbox{\cmp{} collisions.\label{fig:squares:cm+}}[\linewidth]
        {\input{figures/squares-cmplus-collisions-overestimation.tikz}}\\[-4pt]
    \subcaptionbox{ASketch \textftwo{}.\label{fig:squares:asketch}}
        {\input{figures/squares-heavy-key-augmented-filter.tikz}}
    \hfill
    \subcaptionbox{\heavyKeyFTwoNosync{} and \heavyKeyFTwoLagom{}.\label{fig:squares:delegation}}
        {\input{figures/squares-heavy-key-delegation-filter.tikz}}
    \caption{Geometric interpretation of \ftwoest{}.}\label{fig:squares}
\end{minipage}
\end{figure}

\subsection{\cms{} \& Partitioning}
We begin by comparing a `\emph{wide}' CMS with a \emph{partitioned} one, on fixed memory budget.

\noindent
\begin{minipage}{0.66\textwidth}
    \algobox{
        \label{algobox:wide}
        \analysisWide{}: A \cms{} with \sketchHeight{} rows and $\delThreads \times \sketchWidth$ columns. A toy example is shown to the side.
    }
\end{minipage}\hfill%
\begin{minipage}[t]{0.3\textwidth}
    \centering
    \begin{tabular}{|*{4}{m{5mm}|}}
        \hline
        1 &      & 2, 3 & 4 \\
        \hline
        4 & 1, 2 &      & 3 \\
        \hline
    \end{tabular}
\end{minipage}

\noindent\textbf{\textsf{Sequential~}}
\cmp{} (Def.~\ref{defn:cm+}) estimates \ftwo{} from \analysisWide{}.
Consider the example CMS; each number represents a unique key hashed to that counter, and is simultaneously the frequency of that key (i.e., $\frequency{1} = 1$, $\frequency{2} = 2$, etc.).
Colliding keys are separated by commas; CMS will only store their sum.
The true \ftwo{} is $1^2 + 2^2 + 3^2 + 4^2 = 30$.
\cmp{} computes $1^2 + (2+3)^2 + 4^2 = 42$ from the first row and $4^2 + (1+2)^2 + 3^2 = 34$ from the second, returning the smaller estimate.
Overestimation arises from colliding keys.
Fig.~\ref{fig:squares:cm+} illustrates the \cmp{} calculation for the counter containing $(1, 2)$.
The true \ftwo{} contribution of the contained keys is 5, the sum of the plain areas.
\cmp{} calculates $(1+2)^2=9$, the entire square, thus \emph{overestimating by an amount equal to the striped areas}.
The impossibility of underestimating is clear.
Heavy keys induce particularly large extra areas, suggesting a need to treat them specially.\looseness=-1

\noindent\textbf{\textsf{Concurrent~}}
CMS updates and queries can be parallelized by, e.g., atomic fetch-and-add.
In~\cite{rinbergIntermediateValueLinearizability2023}, Rinberg and Keidar show a similar construction to be an IVL implementation of CMS.
On top of this parallel CMS, consider \analysisWideConc{} as a concurrency-aware adaptation of \cmp{} using atomic reads; a similar argument as \cite[Lemma~5.3]{rinbergIntermediateValueLinearizability2023} implies:

\begin{restatable}{lemma}{lemmaConcurrentCMPisIVL}
    \label{lemma:concurrent-cm+-is-ivl}%
    \analysisWideConc{} is an IVL implementation of \cmp{}, preserving \cmp{}'s  \epsdelta{} bounds.
\end{restatable}

For reduced contention relative to \analysisWide{}, the space is now partitioned into \delThreads{} sketches:\looseness=-1

\noindent
\begin{minipage}{0.67\textwidth}
    \algobox{
        \label{algobox:partitioned-cms}
        \analysisPart{}: \delThreads{} partitions, each with a $\sketchHeight \times \sketchWidth$ CMS. Each partition sketches a subdomain of the input. (Toy example on the right).
    }
\end{minipage}\hfill%
\begin{minipage}[t]{0.33\textwidth}
    \centering
        \begin{tabular}{|*{2}{m{5mm}|}}
        \hline
        1 & 2    \\
        \hline
        & 1, 2 \\
        \hline
    \end{tabular}%
    \hspace{0.1em}
    \begin{tabular}{|*{2}{m{5mm}|}}
        \hline
        3, 4 &   \\
        \hline
        4    & 3 \\
        \hline
    \end{tabular}
\end{minipage}

\noindent\textbf{\textsf{Sequential~}}
Partitioning the memory of the example into $\delThreads = 2$ partitions (note each key is only present in one partition),
to estimate \ftwo{} from \analysisPart{}, \analysisPartSeq{} sums a \cmp{} estimate for each partition.
In the example,
$5$ is calculated for the first partition and
$25$ for the second.
Their sum is an \ftwoest{} for the complete \analysisPart{}.

\begin{observation}  \label{obs:partitioned-cm+-improves-accuracy}
    \analysisPartSeq{} can only improve accuracy of \ftwoest{} compared to \cmp{} on \analysisWide{}, as it selects sketch rows with minimum overestimation independently for each partition.
\end{observation}

\noindent\textbf{\textsf{Concurrent~}}
Each partition is assigned a thread to perform updates.
\emph{For now}, assume that each thread only receives tuples for keys of its partition.
Consider \analysisPartConc{} as a parallelization of \analysisPartSeq{} using atomic reads (similar to \analysisWideConc{}) to perform \analysisPartSeq{} concurrently with updates. Similar reasoning as in \autoref{lemma:concurrent-cm+-is-ivl} implies:

\begin{restatable}{lemma}{lemmaPartCMPisIVL}
    \label{lemma:part-cm+-is-ivl}%
    \analysisPartConc{} is an IVL implementation of \analysisPartSeq{}.
\end{restatable}

\subsection{\asketch{} \& Partitioning}  \label{sec:analysis:asketch-and-partitioning}
ASketch~\cite{royAugmentedSketchFaster2016}, §~\ref{sec:background}, introduced filters to track heavy keys more accurately.
In a similar fashion to \analysisPart{}, partitioning can be applied to ASketch while keeping total memory constant:

\algobox{
    \label{algobox:partitioned-asketch}
    \analysisPartASketch{}: \delThreads{} ASketches, each of size $\sketchHeight \times \sketchWidth'$ -- reducing width to fit filters in same total memory.\looseness=-1
}

\noindent\textbf{\textsf{Sequential~}}
Consider \analysisPartASketchSeq{}:
for each partition, \cmp{} is applied to its CMS;
the \ftwo{} contribution of each key \dataKey{} in the filters is $\frequencyEst{\dataKey}^2$, shown in Fig.~\ref{fig:squares:asketch}, where $\frequencyEst{\dataKey} = \augFilterCount{\dataKey}$ (\autoref{algo:asketch:query:heavy-key}).
Note, \frequencyEst{\dataKey{}} contains $o = \augFilterOldCount{\dataKey}$ occurrences which are also in the CMS from times when \dataKey{} did not reside in the filter due to not being heavy enough.
Hence, \cmp{} already included $o^2$ (dotted area in Fig.~\ref{fig:squares:asketch}).
\emph{To avoid double-counting this quantity, $o^2$ is subtracted} (\heavyKeyFTwoNosync{} (Eq.~\ref{eqn:nosync-f2-heavy-keys}) and \heavyKeyFTwoLagom{} (Eq.~\ref{eqn:lagom:f2-heavy-keys})).
Finally, the contributions are summed to get an \ftwoest{}.\looseness=-1

\begin{observation}  \label{obs:more-partitions-more-augfilter-better-accuracy}
    Higher \delThreads{} is beneficial to \analysisPartASketchSeq{}'s accuracy, reducing overestimation through \delThreads{} filters accurately counting heavy keys, while maintaining one-sided error as \cmp{}.\looseness=-1
\end{observation}

\noindent\textbf{\textsf{Concurrent~}}
As in \analysisPartConc{},
each partition is updated by a dedicated thread, (still) with the assumption that input tuples are distributed to the owning partition.
Instead of unsafe scanning, consider \analysisPartASketchConc{} which
obtains an atomic snapshot of each partition-local ASketch (using any synchronization that can guarantee that) and computes an estimate as \analysisPartASketchSeq{}, implying:

\begin{restatable}{lemma}{lemmaPartASketchCMPisIVL}
    \label{lemma:ppacm+-is-ivl}%
    \analysisPartASketchConc{} is an IVL implementation of \analysisPartASketchSeq{}.
\end{restatable}

\subsection{Concurrency Awareness -- Delegation}  \label{sec:analysis:delegation}
We now \emph{waive the simplification of input being distributed to the owning partition}; threads delegate updates to the owner.
We describe and compare three approaches to estimating \ftwo{}.

\algobox{
    \label{algobox:ours}
    \framework{}: Each partition of \analysisPartASketch{} uses \delThreads{} delegation filters to buffer updates, periodically flushed to the owning partition. $\sketchWidth$ is adjusted to maintain the memory budget.
}

\noindent\textbf{\textsf{All DFs~}}
\analysisOursFullSeq{} (Eq.~\ref{eqn:nosync-f2}) extends \analysisPartASketchSeq{} to include all delegation filters when estimating \frequencyEst{\dataKey}, shown as $\sum\!\delFilterCount{}{\dataKey{}}$ in Fig.~\ref{fig:squares:delegation}.
Buffered occurrences of light keys are not considered, as
they by definition do not significantly contribute to \ftwo{}, particularly for skewed streams which we target.
\ftwoestm{\nosync} (§~\ref{sec:overview:elem:nosync}) calculates this as a concurrent query, but lack of synchronization
makes it unsafe (Obs.~\ref{obs:nosync-f2-is-unsafe}).
Further, \analysisOursFullSeq{} scales quadratically in \delThreads{}; all \delThreads{} delegation filters are read from other threads for each of \delFilterSlots{} heavy keys in \delThreads{} ASketch filters.\looseness=-1

\noindent\textbf{\textsf{No DFs~}}
To query more efficiently, we consider outright ignoring delegation filters, which will underestimate \ftwo{}, by applying \analysisPartASketchSeq{} to the \framework{} data structure.
However, the number of buffered, hence ignored, updates per partition is bounded (Obs.~\ref{obs:delegation-filters-buffer-r-updates}).

\begin{restatable}{lemma}{lemmaPartASketchCMPonOursIsRRelaxedIVL}
    \label{lemma:partas-f2-on-ours-is-r-relaxed-ivl}%
    \analysisOursRelaxedConc{} on \analysisOurs{} is an $r$-relaxed IVL implementation of \analysisOursFullSeq{} per partition, where $r = \delThreads \delFilterMaxCount$.
\end{restatable}

\begin{corollary}
    Since IVL is a local property~\cite{rinbergIntermediateValueLinearizability2023}, \analysisOursRelaxedConc{} on \analysisOurs{} is an $r\delThreads$-relaxed IVL implementation of \analysisOursFullSeq{}.
\end{corollary}

Note the worst-case is extremely unlikely to happen, as all $r$ updates would need to be for the same key (i.e. extreme skew), and
filters do not become full simultaneously.

\noindent\textbf{\textsf{Lagom~}}
\label{sec:analysis:delegation:compensated}
To compensate for this relaxation yet maintain its high efficiency, \analysisOursProjSeq{} (§~\ref{sec:overview:lagom}) projects the number of buffered occurrences of heavy keys to replace the exact calculation of $\sum\!\delFilterCount{}{\dataKey{}}$ in Fig.~\ref{fig:squares:delegation} and \heavyKeyFTwoNosync{} (Eq.~\ref{eqn:nosync-f2-heavy-keys}).

\begin{observation}
    \label{obs:lagom-comp-improves-accuracy}
    \heavyKeyFTwoLagom{} (Eq.~\ref{eqn:lagom:f2-heavy-keys}) projection compensates for up to $r/2$ ignored occurrences per heavy key, bringing the query result closer to the actual \ftwo{}, based on the reasoning of the geometric argument and observations \ref{obs:partitioned-cm+-improves-accuracy} and \ref{obs:more-partitions-more-augfilter-better-accuracy}.
\end{observation}

From the above and \autoref{lemma:lagom-is-good}, we have:
\begin{restatable}{lemma}{lemmaF2LagomisIVL}
    \label{lemma:lagom-f2-is-ivl}%
    \analysisOursProjConc{} is an IVL implementation of \analysisOursProjSeq{}.
\end{restatable}

The behavior of \analysisOursProjConc{} within the bounds is data- and execution-dependent and is evaluated in the next section.
Summarizing the properties of \framework{}, we have:
\begin{corollary}
    \framework{} is a composite concurrent sketch data structure for multiple queries:
    (1)~\pq{} (§~\ref{sec:overview:pq}), an IVL implementation of the ASketch point query; %
    (\autoref{lemma:pq-is-ivl});
    (2)~\fonenosync{}, an IVL implementation of exact \fone{} (\autoref{lemma:nosync-f1-is-ivl});
    (3)~\analysisOursProjConc{}, an IVL implementation of \analysisOursProjSeq{} (Obs.~\ref{obs:lagom-comp-improves-accuracy}, \autoref{lemma:lagom-f2-is-ivl}).
    Queries follow the monotonicity properties in \autoref{lemma:monotonicity-of-queries}.\looseness=-1
\end{corollary}

%% file: figures/squares-cmplus-collisions-overestimation.tikz
\begin{tikzpicture}[scale=0.5]
%    \filldraw[thick, densely dotted, pattern=north east lines, pattern color=gray] (0,0) rectangle (3,3);

    \fill[preaction={fill, cyan!40!yellow}, pattern=north east lines, pattern color=olive] (0,1) rectangle (1,3);
    \fill[preaction={fill, cyan!40!yellow}, pattern=north east lines, pattern color=olive] (1,0) rectangle (3,1);
    \draw[semithick, dashed] (0,1) -- (0,3) -- (1,3);
    \draw[semithick, dashed] (1,0) -- (3,0) -- (3,1);

    \filldraw[draw=black, fill=cyan!20] (0,0) rectangle (1,1) node[pos=0.5]{$1^2$};
    \filldraw[draw=black, fill=yellow!50] (1,1) rectangle +(2,2) node[pos=0.5]{$2^2$};
\end{tikzpicture}

%% file: figures/squares-heavy-key-augmented-filter.tikz
\begin{tikzpicture}[scale=0.5]
    \filldraw[pattern=crosshatch dots, pattern color=gray] (0,0) rectangle (1,1) node[pos=0.5]{$o^2$};
    \draw (1,1) rectangle +(2,2) node[pos=0.5]{$\augFilterDrift^2$};
    \draw (1,0) rectangle +(2,1) node[pos=0.5]{$o\!\cdot\!\Delta$};
    \draw (0,1) rectangle +(1,2) node[pos=0.5]{\rotatebox{90}{$o\!\cdot\!\Delta$}};

    \coordinate (B1) at (0,3.3);
    \coordinate (B2) at (3,3.3);
    \draw[|-|] (B1)--(B2) node[above, pos=0.5]{$\frequencyEst{\dataKey}$};

    \coordinate (R1) at (3.4, 0);
    \coordinate (R2) at (3.4, 1);
    \coordinate (R3) at (3.4, 3);
    \draw[|-|] (R1)--(R2) node[right, pos=0.5]{$o$};
    \draw[|-|] (R2)--(R3) node[right, pos=0.5]{$\augFilterDrift$};
\end{tikzpicture}

%% file: figures/squares-heavy-key-delegation-filter.tikz
\begin{tikzpicture}[scale=0.5]
    \filldraw[draw=gray!20, fill=gray!20] (3.0,0) rectangle +(0.9,3.9);
    \filldraw[draw=gray!20, fill=gray!20] (0,3.0) rectangle +(3.9,0.9);
    \draw[densely dashed] (3,0) -- (3.9,0) -- (3.9,3.9) -- (0,3.9) -- (0,3);
%    \draw[densely dotted] (3,3) -- (3.9,3);
%    \draw[densely dotted] (3,3) -- (3,3.9);

    \filldraw[pattern=crosshatch dots, pattern color=gray] (0,0) rectangle (1,1) node[pos=0.5]{$o^2$};
    \draw (1,1) rectangle +(2,2) node[pos=0.5]{$\Delta^2$};

    \filldraw[fill=white] (3,3) rectangle +(0.2,0.2);
    \node at (3.3,3.3)[circle,fill,inner sep=0.3pt]{};
    \node at (3.4,3.4)[circle,fill,inner sep=0.3pt]{};
    \node at (3.5,3.5)[circle,fill,inner sep=0.3pt]{};
    \filldraw[fill=white] (3.6,3.6) rectangle +(0.3,0.3);

    \draw (1,0) rectangle +(2,1) node[pos=0.5]{$o\!\cdot\!\Delta$};
    \draw (0,1) rectangle +(1,2) node[pos=0.5]{\rotatebox{90}{$o\!\cdot\!\Delta$}};

%    \draw[|-|] (3,4.2)--(3.9,4.2) node[above, pos=0.5]{$\sum\!\delFilterCount{}{\dataKey{}}$};
%    \draw[|-|] (4.2,3)--(4.2,3.9) node[right, pos=0.5]{\lagom{} projection};
    \draw[|-|] (4.2,3)--(4.2,3.9) node[right, pos=0.5]{\rotatebox{0}{$\sum\!\delFilterCount{}{\dataKey{}}$}};

%     \coordinate (B1) at (0,-0.3);
%     \coordinate (B2) at (3,-0.3);
%     \draw[|-|] (B1)--(B2) node[below, pos=0.5]{$\mathsf{AF}$};
%
%     \coordinate (R1) at (3.4, 0);
%     \coordinate (R2) at (3.4, 1);
%     \coordinate (R3) at (3.4, 3);
%     \draw[|-|] (R1)--(R2) node[right, pos=0.5]{$\mathsf{Sk}$};
%     \draw[|-|] (R2)--(R3) node[right, pos=0.5]{$\Delta$};
\end{tikzpicture}

%% file: include/06_evaluation.tex
\section{Evaluation}  \label{sec:eval}

\noindent\textbf{\textsf{Baselines~}}
We study \framework{} relative to:
(1)~\swskt{}~\cite{chiosaSKTOnepassMultisketch2021}, the software implementation of SKT which targets FPGA\nobreakdashes-acceleration for high-rate sketching.
SKT uses separate sketches for multiple queries ($F_0$, point queries, and \ftwo{} via HLL, CMS, and \fastagms{}) updated in parallel, but \emph{no concurrent queries} (first merging all thread-local sketches before querying).
This comparison is for insights about
answering mixed queries in a single sketch, from a memory, accuracy and scalability point of view.
(2)~\delsketch{}~\cite{stylianopoulosDelegationSketchParallel2020}, which supports concurrent updates but \emph{with point queries only}.
By comparing, we aim to understand the synchronization overhead for concurrent global queries.
(3)~The elementary \fullsync{} and \nosync{} designs are used to evaluate the %
balance achieved by \lagom{}.
Note that simply adding concurrent queries to \swskt{} would imply effects similar to \nosync{} and \fullsync{} and hence are not studied separately.

\noindent\textbf{\textsf{Datasets and Hardware Platform~}}
From the CAIDA~\cite{CAIDAUCSDAnonymized} network packet traces we extract \qty{18.5}{\mega{}} tuples of headers.
Additionally, synthetic datasets with skew $z$ = \numlist{1;1.5;2;3} (each consisting of \qty{100}{\mega{}} tuples sampled from a domain of size \qty{1}{\mega{}}) are used to explore performance metrics.
All experiments are conducted on a 128-core AMD EPYC 7954 running at \qty{2.25}{\giga\hertz}, without using SMT.
The server runs openSUSE Tumbleweed~20250211.
\framework{} is implemented in \Cpp{}23~\cite{anonymousauthorsLMQSketch}, compiled with GCC~14.2.1 and \texttt{-O3~-march=native}.

\noindent\textbf{\textsf{Metrics~}}
To study the scalability and efficiency of \framework{}, we measure \emph{throughput} (updates per unit-time) at different thread counts, without and with concurrent queries.
To study accuracy in presence of concurrency, we focus on \emph{latency} of global queries,
which has implications on the  IVL-permitted interval;
we propose a methodology to evaluate \emph{accuracy of IVL-associated queries} (cf. §~\ref{sec:analysis}), measuring the difference of returned values relative to the estimated IVL-bounds, giving insight into the synergy of IVL semantics, query latency and freshness.
Finally, we compare the impact of memory budget on accuracy for \ftwoestm{\lagom}, state-of-the-art sketching techniques, and reference methods in §~\ref{sec:analysis}.

\noindent\textbf{\textsf{Parameters~}}
    \emph{Threads, partitions \& memory}\/: we stress-test with high thread counts (\delThreads{} = \numrange{1}{128}).
    Memory budget per partition is constant at \qty{32}{\kibi\byte}: $\sketchHeight\!\times\!\sketchWidth = 8\!\times\!1024$, while reducing \sketchWidth{} when having filters (cf. evaluations in \cite{royAugmentedSketchFaster2016, stylianopoulosDelegationSketchParallel2020}), while keeping sketch rows (hash function computations) fixed.
    \emph{Filter size}\/: \delFilterSlots{} is kept at 16 as in~\cite{stylianopoulosDelegationSketchParallel2020}, delegation filter buffer capacity is bounded at \delFilterMaxCount{} = 1000.
    \emph{Synchronization}: we compare \lagom{} with the aforementioned baselines, including
    two \delsketch{} configurations: the plain one (unbounded $\delFilterMaxCount$) and one with $\delFilterMaxCount\!=\!1000$.
    \emph{Skewness of input data}\/: $z > 1$ for the synthetic data, and $z \approx 1.4$ for the CAIDA data.
    \emph{Frequency of concurrent queries}\/: stress testing with global queries at rates of \qty{1000}{\per\second} and point queries for \qty{0.1}{\percent} of update tuples, following the benchmark in~\cite{stylianopoulosDelegationSketchParallel2020}.

\noindent\textbf{\textsf{Experiments~}}
We begin by comparing the impact of synchronization on update operation's throughput when processing a complete input dataset from system memory (§~\ref{sec:eval:throughput}).
Next, we investigate the accuracy and freshness of global query results, in relation to memory budget and concurrency.
Higher weight lies on the more complicated \ftwo{} queries, \foneest{}'s behavior being studied mainly in terms of query latency and \pq{}'s behavior evaluated in~\cite{stylianopoulosDelegationSketchParallel2020} (§~\ref{sec:eval:mem-acc}, \ref{sec:eval:latency-acc}, \ref{sec:eval:conc-acc}).
For brevity, the results are shown in figures with summary descriptions in the caption, and main takeaways in associated paragraphs. More detailed discussions appear in Appendix~B.\looseness=-1

\subsection{Concurrency and Synchronization}  \label{sec:eval:throughput}

\noindent\textbf{\textsf{Design and Parameters~}}
Each dataset is processed with varying numbers of partitions and threads, \delThreads{}.
First, we measure update throughput with no concurrent queries.
We then repeat the measurement with a constant high load of concurrent queries: \fone{} and \ftwo{} at \qty{1000}{\per\second} each, and point queries at \qty{0.1}{\percent} of updates, for the designs that support them.\looseness=-1

\begin{figure}[ht!]
    \centering
    \includeinkscape[pretex=\scriptsize, width=\linewidth]{insert_throughput_vs_threads_all.pdf_tex}
    \caption{
        Mean update throughput, without (upper row) and with  concurrent queries (lower row, for designs that support them).
        Fluctuations are small and omitted for clarity.
        Note the minimal overhead of \lagom{} for global queries as it matches \delsketch{} with \delFilterMaxCount{}=1k.
        \fullsync{} does not scale with increasing threads or skew.
        With concurrent queries, \lagom{} stays close to \nosync{}, affirming the lightweight-ness of the synchronization design; increasing skew
        improves performance.
    }
    \label{fig:eval:insert-throughput-vs-threads}
    \label{fig:eval:insert-throughput-vs-threads-concurrent-queries}
\end{figure}

\noindent\textbf{\textsf{Takeaways~}}
The results, shown in Fig.~\ref{fig:eval:insert-throughput-vs-threads-concurrent-queries}, imply that \framework{} with \lagom{} is able to maintain state in a way that allows independence by different threads, enabling high processing throughput, yet still imparting the necessary consistency to serve the purpose of sketching for continuously and concurrently answering multiples queries in a streaming setting.\looseness=-1

\subsection{Memory and Accuracy}\label{sec:eval:mem-acc}
We consider a sequential setting, to explore the impact of sketch memory budget on \ftwo{} estimation accuracy for several baselines
and our methods.
\emph{Note}: There are \emph{two main parameters for memory budget}, but with differing impact on accuracy: \emph{per-partition memory} (given by \sketchHeight{} and \sketchWidth{}) and the \emph{total number of partitions \delThreads{}} (also number of threads  when parallelizing).
In the partitioned design, an increase in any of these improves accuracy (§~\ref{sec:analysis}).\looseness=-1

\noindent\textbf{\textsf{Design and Parameters~}}
The methods in the analysis (§~\ref{sec:analysis}) form points of reference; additionally, we compare with \fastagms{}~\cite{cormodeSketchingStreamsNet2005}, used in \swskt{}.
For each method and skew $z$ = \numlist{1;1.5;2}, five synthetic datasets (length 100M, cardinality 1M) are processed under various memory budgets, for \ftwoest{} and the mean absolute percent error (MAPE) relative to the true \ftwo{}, at the end of the execution (i.e., based on the global state for the filter-enabled designs).
For each method in §~\ref{sec:analysis}, we use constant per-partition memory \qty{32}{\kibi\byte} ($\sketchHeight\!\times\!\sketchWidth = 8\!\times\!1024$ with adjustments in presence of filters, as before); total memory increases with \delThreads{}.\looseness=-1

Fig.~\ref{fig:eval:f2-accuracy-sequential} shows the MAPE in log-scale, along the outcome for a single-partition \fastagms{}
at several budgets: $6\!\times\!2^{13}$ (as in the original evaluation of \cite{chiosaSKTOnepassMultisketch2021}),
$8\!\times\!1024$ (same $P=1$ in the partitioned approaches),  and $128\!\times\!8\!\times\!1024$ (identical to the total budget of \framework{} at $\delThreads\!=\!128$).
Increasing parallelism for thread-local designs such as \swskt{} will require \delThreads{} times this memory at runtime,
but the accuracy bounds of queries remain the same as for a single sketch of $1/P$ of the total memory (since the local sketches simply get merged).
Fig.~\ref{fig:eval:memory-scaling} shows memory budget scaling for \swskt{} and \framework{} when the number of partitions/local-sketches follow the number of threads when run in parallel.

\begin{figure}[h]
\begin{minipage}[t]{.62\linewidth}
    \includeinkscape[pretex=\scriptsize, width=\linewidth]{f2_accuracy_sequential.pdf_tex}
    \captionsetup{hypcap=false}
    \captionof{figure}{%
        Non-concurrent MAPE for \textftwo{} estimations at various $z$ and memory budgets, in multiples of 1 partition occupying $\sketchHeight\!\times\!\sketchWidth = 8\!\times\!1024$ counters (\qty{32}{\kibi\byte}).
        Thread-local approaches (\swskt{}) must merge local sketches before querying, and achieve the accuracy of this single sketch, various sizes of which are shown here.
        For such designs, increasing memory budget with additional threads does not improve accuracy.%
    }
    \label{fig:eval:f2-accuracy-sequential}
\end{minipage}\hfill%
\begin{minipage}[t]{.37\linewidth}
    \includeinkscape[pretex=\scriptsize, width=\linewidth]{total_memory_vs_threads.pdf_tex}
    \captionsetup{hypcap=false}
    \captionof{figure}{%
        Total memory scaling linearly with \delThreads{}.
        \framework{} uses \qty{32}{\kibi\byte} per thread\slash{}partition.
        \swskt{}, with $6\!\times\!2^{13}$ counters for its CMS and \fastagms{} as in the evaluation of~\cite{chiosaSKTOnepassMultisketch2021} (but without the HLL for F\textsubscript{0})
        uses \qty{393}{\kilo\byte} per thread.%
    }
    \label{fig:eval:memory-scaling}
\end{minipage}
\end{figure}

\noindent\textbf{\textsf{Takeaways~}}
\analysisPartSeq{} and \analysisPartASketchSeq{}, our stepwise enhancements for \cmp{} in §~\ref{sec:analysis}, improve estimation accuracy.
Unlike for thread-local designs, the partitioned approach allows \framework{} to utilize the increased memory budget for a \emph{two-fold benefit}:
increasing per-partition memory improves (sequential) accuracy in isolation; increasing the number of partitions benefits both accuracy and parallelism.
Although \fastagms{} is one of the most accurate \ftwo{} sketches, its thread-local-oriented design does not see improved accuracy for the same per-partition memory budget, while \framework{} achieves the same or higher accuracy by effectively navigating the memory and concurrency trade-offs.

\subsection{Query Latency and Accuracy}  \label{sec:eval:latency-acc}
Next, we consider concurrent queries.
As described in (G1) (§~\ref{sec:problem}), IVL, while preserving bounds for sketches, cannot fully characterize the final accuracy of the result; query latency plays a key role.
\framework{} addresses this by optimizing query operation latency using partial results at insertion.
We evaluate the latency improvement due to these enhancements.

\noindent\textbf{\textsf{Design and Parameters~}}
We measure the latency of global queries with different values of~\delThreads{}, concurrent with updates.
After \qty{100}{\milli\second} of warmup to populate delegation filters, queries are performed at a rate of \qty{1000}{\per\second}.
For \fone{} queries, we focus on the \fonenosync{} IVL design (§~\ref{sec:overview:elem:nosync}).
For \ftwo{}, the more complex query, we benchmark the synchronization designs.

\begin{figure}[h]
\begin{minipage}[t]{.38\linewidth}
    \includeinkscape[pretex=\scriptsize, width=\linewidth]{f1_scanner_latency_vs_threads.pdf_tex}
    \captionsetup{hypcap=false}
    \captionof{figure}{%
        Latency of \fonenosync{} queries concurrent with updates scales linearly with~\delThreads{}.
        No impact from skew.
    }
    \label{fig:eval:f1-scan-latency}
\end{minipage}\hfill%
\begin{minipage}[t]{.6\linewidth}
    \includeinkscape[pretex=\scriptsize, width=\linewidth]{f2_scanner_latency_vs_threads.pdf_tex}
    \captionsetup{hypcap=false}
    \captionof{figure}{%
        Latencies of 100 \textftwo{} queries concurrent with updates.
        \lagom{} is 2-3 order of magnitude faster than other designs, which struggle with high thread counts.
    }
    \label{fig:eval:f2-scan-latency}
\end{minipage}
\end{figure}

\noindent\textbf{\textsf{Takeaways~}}
\fonenosync{} query duration ranges from \qty{100}{\nano\second} to \qty{10}{\micro\second} at high thread counts.
Our optimizations in \ftwoestm{\lagom} yield significant improvements in query latency, which stays below \qty{100}{\micro\second}.
Such short latencies suggest the suitability of our queries for accurate estimation under IVL, due to reduced number of overlapping updates, evaluated next.

\subsection{Concurrency and Accuracy}\label{sec:eval:conc-acc}
Building on insights from measuring concurrent query latencies, we now evaluate the impact on accuracy.
Query duration in conjunction with update throughput determine the IVL interval size.
Although IVL does not exactly target accuracy guarantees, to complement, we explore the admissible freshness
of this interval.

\noindent\textbf{\textsf{Design and Parameters~}}
We compare the return value of concurrent \ftwoestm{\lagom} with the IVL-permitted interval.
We adopt the following methodology to determine the relative error induced by concurrent executions, without interfering with execution patterns, e.g., as an approach based on stopping and starting updater threads to record measurements would do.
\\\algobox{%
    \textbf{\textsf{Methodology to determine bounds of return interval of concurrent query $Q$}}:
    We reconstruct ideal query return values at \qstart{} and \qend{}.
    A threshold \fone{}-value $T$ is selected, beyond the warmup period of the data structure, as a trigger point for performing $Q$ which overlaps an arbitrary number of updates $n$.
    To reconstruct \qstart{}, updaters are stopped when \fone{} reaches $T$ and a sequential query is performed.
    In a new execution, reaching $\fone{} = T$ instead triggers $Q$ concurrently.
    Upon the completion of $Q$, updater threads are stopped and \qend{} is recorded in a sequential setting.
}
We set $T$ = 10M tuples and perform 50 repetitions for each dataset (CAIDA and synthetic with $z$ = \numlist{1;1.5;2}) and increasing \delThreads{}.
We study \ftwoestm{\nosync{}} and \ftwoestm{\lagom} using the same methodology to determine the effect of latency and weaker semantics on accuracy.

\begin{figure}[htb]
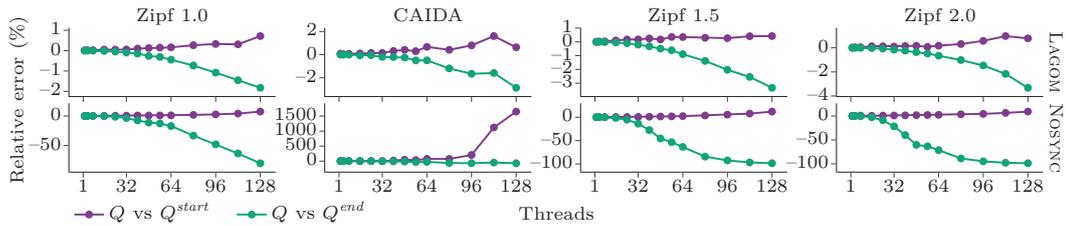

    \centering
    \includeinkscape[pretex=\scriptsize, width=\linewidth]{f2_accuracy_concurrent-all.pdf_tex}
    \captionsetup{hypcap=false}
    \caption{%
        Comparison of query return value against bounds of IVL-permitted interval.
        For both designs, \qop{} observes more updates than \qstart{} (which is what \ftwoestm{\fullsync} would return) but misses some updates observed by \qend{}.
        \lagom{} consistently yields narrower return value intervals compared to \nosync{}, which may miss many updates due to its long operation latency.
    }
    \label{fig:eval:f2-accuracy-concurrent}
\end{figure}

\noindent\textbf{\textsf{Takeaways~}}
The concurrent \ftwo{} queries can observe overlapping updates (towards improving  \fullsync{}'s freshness).
However, the interval of return values for \nosync{} is arbitrarily large, while \lagom{} (with  agile calculation and IVL guarantees) imparts a seriously smaller uncertainty, particularly at low--intermediate \delThreads{}, demonstrating that \lagom{}'s properties preserve, and are useful for, accuracy.

%% file: include/07_relatedwork.tex
\section{Other Related Work}  \label{sec:relatedwork}

Rinberg et al. in~\cite{rinbergFastConcurrentData2022} present a snapshot-based methodology for sketch \emph{querying concurrently with updates}, building on a thread-local design with limited local buffering and a global propagator which merges local sketches into a global one when full.
Concurrent queries require a  snapshot of this global sketch,
which gets more costly with the size of the state, also leading to a tradeoff with respect to accuracy.
Using a partitioning design as in \framework{} allows efficient queries taking snapshots per partition,
while supporting distributed state and fine-tuning of freshness.
Concurrent queries and updates are also targeted in~\cite{eliaszadaQuancurrentConcurrentQuantiles2023} for estimating quantiles, although not using the generic framework of \cite{rinbergFastConcurrentData2022} due to risk of sequential bottlenecks.\looseness =-1

Several recent orthogonal works such as Hydra~\cite{manousisEnablingEfficientGeneral2022} and OmniSketch~\cite{punterOmniSketchEfficientMultiDimensional2023} explore using sketches for \emph{multiple querying} of multidimensional data streams.
These works are not targeting queries concurrent with updates, though, instead using an online sketching phase followed by an offline querying phase, during which the sketch is no longer updated.

\emph{Universal Sketching} appears in~\cite{liuOneSketchRule2016, manousisEnablingEfficientGeneral2022} as a useful technique to estimate (a subset of) a class of functions from a single sketch.
As its core functions associate with the ones in \framework{}, the latter can be a possible candidate for supporting concurrency in it.

%% file: include/08_conclusions.tex
\section{Conclusions}  \label{sec:conclusion}

We presented \framework{}, a concurrent, multi-query sketch in a single, low memory-footprint object, with explicit concurrency semantics that lead to predictably high accuracy even with very demanding streams.
Our analytical and detailed empirical insights show that
(1)~having a single data structure balances multiple targets: accuracy (Fig.~\ref{fig:eval:f2-accuracy-sequential}), timeliness (Fig.~\ref{fig:eval:f1-scan-latency}, \ref{fig:eval:f2-scan-latency}), memory footprint (Fig.~\ref{fig:eval:memory-scaling}), freshness (Fig.~\ref{fig:eval:f2-accuracy-concurrent}), concurrency, and throughput (Fig.~\ref{fig:eval:insert-throughput-vs-threads});
(2)~besides the memory budget, the way that the memory is used is catalytic for the achievable concurrency and accuracy, both for local and global queries;
(3)~employing IVL, in conjunction with low query latency and concurrency-aware compensation, allows accuracy aligning with that of the sequential sketches, even under very high-rate, skewed streams.

\framework{} can be a useful component in systems such as Redis, Apache Druid/Spark\slash{}DataSketches and more.
It is extensible to answer other queries, e.g. top-k elements, and possibly support wavelets and quantiles.
\framework{}'s set of queries associate with the ones required for universal sketching and hence make it a possible candidate for such a concurrent construction.\looseness=-1

%% file: include/a_proofs.tex
\section{Proofs}  \label{appendix:proofs}

Due to space limitations, arguments that support the claims in the main part of the paper are presented in more detail in this appendix.

\subsection{From §~\ref{sec:overview}}

\lemmaPQisIVL*{}
\begin{proof}[Proof sketch]
    While \delThreadI{\delOwner{\dataKey{}}} executes \pq{}, other threads may concurrently update relevant delegation filters.
    However, only updates overlapping \pq{} can be missed; updates completed before \pq{} are reflected in the returned \frequencyEst{\dataKey{}}~\cite[Claim~2]{stylianopoulosDelegationSketchParallel2020}.
    Double-counting of updates is not possible~\cite[Claim~3]{stylianopoulosDelegationSketchParallel2020}, as this would mean \pq{} observed an update both while it is buffered in a delegation filter, as well as when it has been flushed to the partition-local ASketch.
    This would require a flush of the delegation filter, which can only be performed by the same thread as \pq{} itself, and hence cannot overlap \pq{}.
    Therefore, the maximal return value is the one for a linearization which observes all overlapping updates.
\end{proof}

\lemmaNosyncFOneIsIVL*{}
\begin{proof}[Proof sketch]
    The partial results form jointly a shared counter, where each thread maintains a local counter of completed updates.
    The query, performing one atomic read per partition, can neither double-count updates not omit completed ones.
\end{proof}

\obsNosyncFTwoIsUnsafe*{}
This can occur when $Q$ overlaps with processing delegated updates (Alg.~\ref{algo:delsketch:ppi}), which non-atomically moves occurrences of keys from delegation filters to the ASketch of the owning partition.
Updates can be missed when $Q$ scans a sketch \delThreadSketch{\delThreadIIndex{}} (\autoref{algo:nosync:f2:get-sketch-cmplus}), whereupon a filter \delThreadDelFilter{\delThreadIIIndex}{\delThreadIIndex} is handed over to \delThreadI{\delThreadIIndex} and flushed (Alg.~\ref{algo:delsketch:ppi}),
followed by $Q$ reading the now-empty filter (\autoref{algo:nosync:f2:heavy-key-foreach-df}), thus missing all the concurrently flushed updates, regardless of whether they were completed before $Q$ began.
Or, if $Q$ overlaps a flush of \delThreadDelFilter{\delThreadIIIndex}{\delThreadIIndex}, updates may be seen both in the ASketch of \delThreadI{\delThreadIIndex} and in \delThreadDelFilter{\delThreadIIIndex}{\delThreadIIndex} before the latter is cleared.\footnote{Similarly, there can there be omissions and double-counting if we swap the relative order of scanning of sketches and delegation filters.}

\lemmaLagomAtomicSnapshotAndNoDeadlocks*{}
\begin{proof}
    \emph{Atomic per-partition snapshot --} 
    When a full delegation filter is flushed, buffered updates move to the owning partition's ASketch and associated partial results are updated (Alg.~\ref{algo:asketch:insert}).
    As this operation is not atomic, the query synchronizes with it to avoid mis-calculations, such as double-counting updates or omitting completed ones:
    \begin{itemize}
        \item A flush begins by incrementing \versionOne{} and increments \versionTwo{} upon completion (\autoref{algo:lagom:ppi:ts1} and \ref{algo:lagom:ppi:ts2}).
        A query reads these in reverse order (\versionTwo{} then \versionOne{}) to detect concurrent modification of the data it read, and the scan is retried is they do not match (condition on \autoref{algo:lagom:f2:versions-match}).
        \item To prevent unbounded retrying in the (unlikely) event that repeated flushes occur in a query's duration, the query flags the partition it is scanning (\autoref{algo:lagom:f2:set-flag}).
        If the flush finds the flag set (\autoref{algo:lagom:ppi:flag-check}), it waits until the query has cleared the flag (\autoref{algo:lagom:f2:clear-flag}).
        \item If the flag is not set, the flush can proceed to increment \versionOne{} (\autoref{algo:lagom:ppi:ts1}) and begin updating the ASketch contents.
        If a query now sets the flag, it will either find the version numbers match and knows that a consistent view was obtained, or detect a mismatch, indicating the observed state may be inconsistent, and retry the scan (\autoref{algo:lagom:f2:versions-match}).
        The query can be blocked by \emph{at most one} concurrent filter flush, as subsequent ones will be stalled by the flag.
    \end{itemize}
    The synchronization design performs a double-collect of the version numbers (\autoref{algo:lagom:f2:versions-match}) which indicate concurrent modification of the ASketch of the partition being scanned.
    As a scan gets an image of a partition's partial results without interferences from filter flushes, it sees a linearizable view of that information in the partition.

    \emph{No deadlock --} 
    Deadlock entails one or more updater threads and the \ftwo{} query thread being unable to proceed because they are waiting for one another.
    The query algorithm interacts with only one partition at a time; all remaining updater threads for other partitions are not involved and continue executing independently.
    If the query has set the flag for updater thread \delThreadI{\delThreadIIndex} (\autoref{algo:lagom:f2:set-flag}), \delThreadI{\delThreadIIndex} will not proceed past \autoref{algo:lagom:ppi:flag-check}.
    Within bounded steps, the query will complete its scan of partition \delThreadIIndex{}, as \delThreadI{\delThreadIIndex} is currently blocked and cannot update the version numbers to mismatch, and ultimately reset the flag (\autoref{algo:lagom:f2:clear-flag}), thus unblocking \delThreadI{\delThreadIIndex}.
    Similarly, if the query is attempting obtain a consistent scan while \delThreadI{\delThreadIIIndex} is performing a filter flush operation and has incremented \versionOne{} -- causing the query to retry --, from our system properties, \delThreadI{\delThreadIIndex} is guaranteed to make progress and complete the flush within bounded time, finally incrementing \versionTwo{} to match.
    Again, the query will be blocked by \emph{at most one} concurrent filter flush, as subsequent ones will be stalled by the flag.
\end{proof}

\lemmaMonotonicityOfLagomQueries*{}

\begin{proof}
    If $Q_1$ and $Q_2$ are of the same type or if $\subop{U}{Q_2} \precedes \subop{U}{Q_1}$ according to program order, then monotonicity of scans is immediate.
    There are three remaining cases where it is possible that $\subop{U}{Q_1} \precedes Q_1 \precedes Q_2 \precedes \subop{U}{Q_2}$ when $U$ overlaps both queries:
    \begin{description}
        \item[\textpq{} \precedes{} \textfone{}]
        At most one update operation per updater thread can overlap both $Q_1$ and $Q_2$.
        Therefore, a \pq{} is limited to observing at most \delThreads{} updates that are not yet visible to a subsequent \fone{} query.

        \item[\textpq{} \precedes{} \textftwo{}]
        Since $Q_1$ is a local query, hence only observes one partition, at most $r = \delThreads \delFilterMaxCount$ updates buffered in delegation filters for this partition may be observed by $Q_1$ but not by $Q_2$ (\autoref{obs:delegation-filters-buffer-r-updates}).

        \item[\textfone{} \precedes{} \textftwo{}]
        Similar to the previous case; however, $Q_1$ is now a global query and may in the worst case observe up to $r = \delThreads \delFilterMaxCount$ buffered updates \emph{per partition} not yet visible to $Q_2$.
        Hence, the queries may deviate from each other by up to $r \delThreads$ updates, but no more.\qedhere
    \end{description}
\end{proof}

\subsection{From §~\ref{sec:analysis}}

\lemmaConcurrentCMPisIVL*{}
\begin{proof}[Proof sketch]
    Following a similar argument as \cite[Lemma~5.3]{rinbergIntermediateValueLinearizability2023}; %
    as updates only increment counters, each counter read by a \analysisWideConc{} query $Q$ will return a value at least as large as the value at the start of $Q$, and no larger than the value at the end of $Q$.
    Transitively, this holds also for the sum of squared values of said counters as computed by \analysisWideConc{}.
    The returned \ftwoest{} will be the minimum of these per-row results, and it cannot be lower than the least per-row result at the start of the query, or larger than the least per-row result at the end of the query,
    as required for IVL of monotonically increasing quantities (\autoref{obs:monotonic-ivl-is-bounded-between-start-and-end}).
\end{proof}

\lemmaPartCMPisIVL*{}
\begin{proof}[Proof sketch]
    Similar reasoning as for \autoref{lemma:concurrent-cm+-is-ivl}.
\end{proof}

\lemmaPartASketchCMPisIVL*{}
\begin{proof}[Proof sketch]
    Similar to earlier lemmas.
    By virtue of being atomic, the snapshots for each partition will linearize with updates by the corresponding updater thread.
    Since counter values read by the query are non-decreasing, the return value is bounded between an ideal return value at the start of the query (observing all preceding updates) and an ideal return value at the end of the query (additionally observing all concurrent updates).
\end{proof}

\lemmaPartASketchCMPonOursIsRRelaxedIVL*{}
\begin{proof}[Proof sketch]
    For a partition, up to $r = \delThreads \delFilterMaxCount$ updates buffered in delegation filters (\autoref{obs:delegation-filters-buffer-r-updates}) may be missing from the atomic snapshot taken by \analysisOursRelaxedConc{}.
    Hence, the return value is bounded between the value at query start excluding the $r$ buffered updates, and the value at the end, including all $r$ buffered updates.
\end{proof}

%% file: include/b_evaluation.tex
\section{Evaluation \& Discussion}

Due to space limitations, the discussion supporting the main takeaways of the empirical study is presented in more detail in this appendix.

\subsection{Concurrency and Synchronization (§~\ref{sec:eval:throughput}) -- Results}
Fig.~\ref{fig:eval:insert-throughput-vs-threads} shows the mean rate of update operations for sketching the complete input data sets.
\fullsync{} cannot scale and performance worsens with more threads as contention around the global RW lock increases.
\nosync{} scales similarly to the plain \delsketch{} (neither of them support consistent global queries).
As expected, \swskt{} throughput grows linearly with the number of threads, as threads process updates entirely independently.
However, concurrent queries cannot be supported.
With increasing skewness, a clear upwards trend in throughput of the delegation design is seen, as delegation filters are able to buffer more updates locally, reducing inter-thread communication.

As expected, limiting local buffering by the parameter \delFilterMaxCount{} impacts update throughput (\delsketch{} with \delFilterMaxCount{} = \num{1000}); still, throughput remains better or similar to the purely thread-local design of \swskt{} without any communication.
\lagom{} exhibits very similar scaling as \delsketch{} with \delFilterMaxCount{} = \num{1000}, processing approximately 1.5~billion updates/second on real-world data, demonstrating the very low overhead of its synchronization design for consistent global queries.
The growth in update throughput of \lagom{} is not linear, initially growing faster than a thread-local design (\swskt{}), and eventually plateauing.
Outperforming \swskt{} is expected, as updates to filters are generally faster (touching 1 counter) than performing entire sketch updates (performing \sketchHeight{} hash computation, touching equally many counters).
Eventually, particularly with high skew where a single heavy key accounts for a large proportion of the input stream, threads frequently hand over delegation filters to the owner of this key, inducing the plateau.

On the lower row, Fig.~\ref{fig:eval:insert-throughput-vs-threads-concurrent-queries} shows the mean rate of update operations with concurrent queries for designs which support them.
\fullsync{} shows similar scaling to before.
\lagom{} achieves very similar performance in presence of concurrent queries, with 1.5~billion updates/second on the real-world CAIDA data and exceeding 2~billion for skew levels \numlist{1.5;2}.
Decreases in update throughput are explained by the fact that updater threads are responsible for serving point queries alongside their update workload.
\nosync{} clearly shows this effect as global queries have no impact in this design, but the overhead of point query work leads to a reduction in peak throughput by around \qty{50}{\percent} compared to Fig.~\ref{fig:eval:insert-throughput-vs-threads} at \qty{0.1}{\percent} point queries (exactly reproducing the result in \cite{stylianopoulosDelegationSketchParallel2020}).
\lagom{}, on the other hand, does not slow down compared to before, despite performing the necessary additional synchronization for global queries.
As motivated regarding the choice of baselines, simply adding concurrent queries to \swskt{} would lead to behavior similar to \nosync{} and \fullsync{}, hence not shown here explicitly.

\subsection{Memory and Accuracy (§~\ref{sec:eval:mem-acc}) -- Results}
Fig.~\ref{fig:eval:f2-accuracy-sequential} shows the accuracy of various \ftwo{} methods, without concurrent updates when a query executes.
As discussed in §~\ref{sec:analysis}, \analysisPartSeq{} and \analysisPartASketchSeq{}, our enhanced versions of \cmp{} for more efficient data structures, improve estimation accuracy.
Further, for partitioning-based approaches, increasing the memory budget by increasing the number of partitions leads to improved accuracy, as more memory is available for (1) avoiding hash collisions (\autoref{obs:partitioned-cm+-improves-accuracy}) and (2) accurate tracking of heavy keys in ASketch filters (\autoref{obs:more-partitions-more-augfilter-better-accuracy}).
On the other hand, thread-local approaches (such as  in \swskt{}) which utilize additional memory for parallelizing updates but require merging all thread-local sketches for querying, see accuracy equivalent to the memory of a single sketch.
Although \fastagms{}, used in \swskt{} is one of the most accurate techniques for \ftwo{} estimation for arbitrary skewness, a thread-local design does not see improved accuracy for the same memory budget, while our partition-targeting methods, navigating memory and concurrency trade-off challenges, achieve higher accuracy.
Of course, to be fair, one should observe that \swskt{} was not designed with concurrent queries as target.

\subsection{Latency and Accuracy (§~\ref{sec:eval:latency-acc}) -- Results}
Fig.~\ref{fig:eval:f1-scan-latency} shows \fonenosync{} query latency, which grows linearly with the number of threads and partitions --- as expected from Eq.~\ref{eqn:nosync-f1} --- and is not impacted by skew.
Query duration is small, ranging from \qty{100}{\nano\second} to \qty{10}{\micro\second} at high thread counts, which implies a low number of overlapping updates.

Similarly, Fig.~\ref{fig:eval:f2-scan-latency} shows the distribution of \ftwo{} query durations for various synchronization designs.
The operational complexity of \ftwoestm{\nosync{}} scales with the number of delegation filters and ASketch filters, as for each slot in ASketch filters, all delegation filters of that partition are read (\heavyKeyFTwoNosync{}, Eq.~\ref{eqn:nosync-f2-heavy-keys}).
\ftwoestm{\fullsync{}} tends to perform slightly better at larger thread counts, as the query has priority over updater threads when acquiring the global lock, and reads the complete memory of the data structure `only' once.
Our \ftwoestm{\lagom{}} is significantly faster, both absolutely and comparatively, at less than \qty{100}{\micro\second}, \numrange{2}{3} orders of magnitude faster than \ftwoestm{\fullsync{}} and \ftwoestm{\nosync{}}, regardless of skew.
We see the impact of our latency optimization; \ftwoestm{\lagom{}} approximates the calculation performed by \ftwoestm{\nosync{}} (\analysisOursProjSeq{} vs \analysisOursFullSeq{}), but is significantly faster.

\subsection{Concurrency and Accuracy (§~\ref{sec:eval:conc-acc}) -- Results}
Fig.~\ref{fig:eval:f2-accuracy-concurrent} shows how query return values relate to IVL interval boundaries for \ftwoestm{\lagom} and \ftwoestm{\nosync}.
In all cases, the concurrent query $Q$ observes more updates than \qstart{} --- which is what \ftwoestm{\fullsync} would return --- but ignores some updates that \qend{} observes, as expected.
Thus, the truly concurrent \ftwoest{} query results are more fresh than \fullsync{} synchronization.
However, the size of the interval for \nosync{} is large compared to \lagom{}, due to its significantly longer query latency (seen in Fig.~\ref{fig:eval:f2-scan-latency}) and the weakness of its semantics (described in Obs.~\ref{obs:nosync-f2-is-unsafe}).
The real-world CAIDA dataset exhibits a more noisy behavior (CAIDA datasets commonly contain anomalies) than the synthetic datasets, reflected in the differing shape of the traces here, particularly for \nosync{}; nonetheless, the behavior of our queries is consistent: i.e.,
\lagom{}, imparts a much smaller IVL-interval on the return value of $Q$, particularly at low and intermediate thread numbers, demonstrating that our lightweight synchronization preserves accuracy of results close to the sequential expectation.

\subsection{Overall Takeaways}
While IVL, as a useful correctness criterion, allows reasoning about semantics of concurrent queries and can preserve \epsdelta{} bounds of sketches, it cannot alone fully characterize the accuracy of results.
Our study shows that \ftwoestm{\lagom}'s compensation scheme is very accurate in a sequential setting, it utilizes available memory for improved parallelism and accuracy in a concurrent one, where efficient synchronization permits low query latency, implying freshness of returned results.
These observations illustrate the challenges tackled in this work.